\newif\ifdraftmode
\newif\iftodominormode

\todominormodetrue

\draftmodefalse
\InputIfFileExists{style.tex}{}{}

\ifdraftmode
  \documentclass[12pt,american,english,intlimits]{article}
\else
  \documentclass[10pt,american,english,twocolumn,twoside,headsepline,footsepline]{scrartcl}
\fi
\newcommand{\mytitle}{Six networks on a universal neuromorphic computing substrate}
\newcommand{\myrunningtitle}{A universal neuromorphic computing substrate}
\title{\mytitle}
\newcommand{\myauthor}{Pfeil et al.}

\long\def\symbolfootnote[#1]#2{\begingroup%
\def\thefootnote{\fnsymbol{footnote}}\footnote[#1]{#2}\endgroup}

\newcommand{\kip}{\textsuperscript{1}}
\newcommand{\schmuker}{\textsuperscript{2,3}}
\author{
Thomas Pfeil\footnotemark[1]~\footnotemark[2]~\kip
\and Andreas Gr\"{u}bl\footnotemark[2]~\kip
\and Sebastian Jeltsch\footnotemark[2]~\kip
\and Eric M\"{u}ller\footnotemark[2]~\kip
\and Paul M\"{u}ller\footnotemark[2]~\kip
\and Mihai A. Petrovici\footnotemark[2]~\kip
\and Michael Schmuker\footnotemark[2]~\schmuker
\and Daniel Br\"{u}derle\kip
\and Johannes Schemmel\kip
\and Karlheinz Meier\kip
}

\usepackage[T1]{fontenc}
\usepackage[utf8]{inputenc}
\usepackage{prettyref} 
\usepackage{float}     
\usepackage{multirow}
\usepackage{amssymb}
\usepackage{amsmath}
\usepackage{graphicx}
\usepackage{setspace}  
\usepackage[authoryear]{natbib}
\usepackage{psfrag}
\usepackage[disable]{todonotes}
\usepackage{booktabs}
\usepackage{siunitx}
\usepackage{printlen}

\usepackage{tabularx}
\newcolumntype{L}[1]{>{\raggedright\arraybackslash}p{#1}} 
\newcolumntype{C}[1]{>{\centering\arraybackslash}p{#1}}   
\newcolumntype{R}[1]{>{\raggedleft\arraybackslash}p{#1}}  

\ifdraftmode
  \usepackage[
  colorlinks=true,linkcolor=black,citecolor=black,urlcolor=black,filecolor=black,
  pdffitwindow,
  pdftitle={\mytitle{}},
  pdfauthor={},
  pdfsubject={},
  pdfkeywords={},
  ]{hyperref}
\else
  \usepackage[
  colorlinks=true,linkcolor=gray,citecolor=gray,urlcolor=gray,filecolor=gray,
  pdffitwindow,
  pdftitle={\mytitle{}},
  pdfauthor={},
  pdfsubject={},
  pdfkeywords={},
  draft
  ]{hyperref}
\fi

\usepackage{fixltx2e}
\usepackage[a4paper]{geometry}
\ifdraftmode
\else
  \usepackage{dblfloatfix}
  \usepackage{sectsty}
  \allsectionsfont{\sf}
\fi

\usepackage{fancyhdr}
\makeatletter

\ifdraftmode
  \fontfamily{ptm}
  \fontseries{m}
  \fontshape{n}
  
\fi

\newcommand{\strong}[1]{\@strong{#1}}
\newcommand{\@@strong}[1]{\textbf{\let\@strong\@@@strong#1}}
\newcommand{\@@@strong}[1]{\textnormal{\let\@strong\@@strong#1}}
\let\@strong\@@strong

\newcommand{\spikey}{\emph{Spikey}}
\newcommand{\network}{Network Topology}
\newcommand{\hardware}{Hardware Emulation}

\newcommand{\swpackage}[1]{\texttt{#1}}

\newcommand{\mrmsub}[2]{#1_{\text{#2}}}
\newcommand{\itsub}[2]{#1_{#2}}
\newcommand{\paramtype}[2]{\mathsf{#1_{#2}}}

\newcommand{\smalltodo}[2][]
{\todo[caption={#2}, #1]
{\begin{spacing}{0.5}#2\end{spacing}}}
\newcommand{\verysmalltodo}[2][]
{\smalltodo[size=\tiny, #1]{#2}}

\iftodominormode
  \newcommand{\smalltodominor}[2][]{\smalltodo[color=yellow,#1]{#2}}
  \newcommand{\verysmalltodominor}[2][]{\verysmalltodo[color=yellow,#1]{#2}}
\else
  \newcommand{\smalltodominor}[2][]{}
  \newcommand{\verysmalltodominor}[2][]{}
\fi

\newrefformat{cap}{\hyperref[#1]{Figure~\ref{#1}}}
\newrefformat{fig}{\hyperref[#1]{Figure~\ref{#1}}}
\newrefformat{sec}{\hyperref[#1]{Section~\ref{#1}}}
\newrefformat{sub}{\hyperref[#1]{Section~\ref{#1}}}
\newrefformat{tab}{\hyperref[#1]{Table \ref{#1}}}
\newrefformat{eq}{\hyperref[#1]{Equation~\ref{#1}}}

\fancypagestyle{plain} 
{

\fancyhf{}
\fancyfoot[L]{}
\fancyfoot[C]{}
\fancyfoot[R]{}
}

\renewenvironment{abstract}
{\noindent{\normalfont\large\textbf{Abstract}}%
\par\vspace{0.5\baselineskip}\noindent}
{\par}

\renewcommand{\@seccntformat}[1]{%
\csname the#1\endcsname\hspace{0.5em}}

\ifdraftmode
  \renewcommand{\section}{\@startsection
  {section}%
  {1}%
  {0mm}%
  {-\baselineskip}%
  {0.5\baselineskip}%
  {\normalfont\large\bfseries}}

  \renewcommand{\subsection}{\@startsection
  {subsection}%
  {1}%
  {0mm}%
  {-\baselineskip}%
  {0.5\baselineskip}%
  {\normalfont\bfseries}}

  \renewcommand{\subsubsection}{\@startsection
  {subsubsection}%
  {2}%
  {1em}%
  {-\baselineskip}%
  {-\fontdimen2\font plus -\fontdimen3\font minus -\fontdimen4\font}%
  {\normalfont\bfseries}}

\else
  \renewcommand\@seccntformat[1]{\csname the#1\endcsname.\quad}
  \renewcommand*\section{\@startsection{section}{1}{\z@}%
  {-0.5ex \@plus -1ex \@minus -.2ex}%
  {0.5ex \@plus.2ex}%
  {\raggedsection\normalsize\bfseries\sffamily\color{blue}\nobreak\MakeUppercase}%
  }
  \renewcommand*\subsection{\@startsection{subsection}{2}{\z@}%
  {\baselineskip}%
  {0.25ex \@plus .2ex}%
  {\raggedsection\small\bfseries\sffamily\nobreak\MakeUppercase}%
  }
  \renewcommand*\subsubsection{\@startsection{subsubsection}{3}{\z@}%
  {\baselineskip}%
  {0.25ex \@plus .2ex}%
  {\raggedsection\small\bfseries\sffamily\itshape\nobreak}%
  }
  \setcapindent{0pt}
  \setkomafont{captionlabel}{\bfseries\sffamily}
  \setkomafont{caption}{\small\sffamily}
  \renewcommand{\fnum@figure}{\textbf{FIGURE~\thefigure}}
  \renewcommand{\fnum@table}{\textbf{TABLE~\thetable}}
\fi

\usepackage{babel} 

\@ifundefined{showcaptionsetup}{}{%
 \PassOptionsToPackage{caption=false}{subfig}}
\usepackage{subfig}
\makeatother

\usepackage{babel}

\global\long\def\vmem{\mrmsub{V}{m}}
\global\long\def\cmem{\mrmsub{C}{m}}
\global\long\def\gl{\mrmsub{g}{l}}
\global\long\def\taum{\mrmsub{\tau}{m}}
\global\long\def\el{\mrmsub{E}{l}}
\global\long\def\eexc{\mrmsub{E}{exc}}
\global\long\def\einh{\mrmsub{E}{inh}}
\global\long\def\vreset{\mrmsub{V}{reset}}
\global\long\def\vth{\mrmsub{V}{th}}
\global\long\def\tauref{\mrmsub{\tau}{refrac}}
\global\long\def\weight{w}
\global\long\def\tausyn{\mrmsub{\tau}{s}}       
\global\long\def\taustdp{\mrmsub{\tau}{STDP}}   
\global\long\def\aplusminus{\mrmsub{A}{+/-}}    
\global\long\def\taurec{\mrmsub{\tau}{rec}}     
\global\long\def\taufacil{\mrmsub{\tau}{facil}} 
\global\long\def\gimax{\itsub{g}{i}^\text{max}}
\global\long\def\trise{\mrmsub{t}{rise}}
\global\long\def\tfall{\mrmsub{t}{fall}}

\global\long\def\cc{\mrmsub{C}{C}}
\global\long\def\cv{\mrmsub{C}{V}}

\begin{document}

\newlength{\fullfigwidth}
\newlength{\columnfigwidth}
\ifdraftmode
  \setlength{\fullfigwidth}{18cm}
  \setlength{\columnfigwidth}{85mm}
\else
  \setlength{\fullfigwidth}{\textwidth}
  \setlength{\columnfigwidth}{\linewidth}
\fi

\begin{titlepage}\thispagestyle{empty}\pdfbookmark[1]{Title}{TitlePage}

\ifdraftmode
\begin{center}
\textbf{\LARGE \mytitle{}}
\par\end{center}{\LARGE \par}

\begin{center}
\textbf{\large Thomas Pfeil{*}$^{\dagger 1}$, Andreas Gr{\"u}bl$^{\dagger 1}$,
Sebastian Jeltsch$^{\dagger 1}$, Eric M{\"u}ller$^{\dagger 1}$, Paul M{\"u}ller$^{\dagger 1}$, Mihai A. Petrovici$^{\dagger 1}$, Michael Schmuker$^{\dagger 2,3}$,
Daniel Br{\"u}derle$^{1}$, Johannes Schemmel$^{1}$, Karlheinz Meier$^{1}$}
\par\end{center}{\large \par}
\else
\maketitle
\fi

\vfill{}

\begin{flushleft}
$^{1}$\parbox[t]{10cm}{Kirchhoff-Institute for Physics\\
Universit{\"a}t Heidelberg\\
Heidelberg, Germany\raisebox{11cm}[0cm][0cm]{\hspace*{5cm}\textbf{\ }}}\\[6mm]
\par\end{flushleft}

\begin{flushleft}
$^{2}$\parbox[t]{10cm}{Neuroinformatics \& Theoretical Neuroscience\\
Freie Universit\"at Berlin\\
Berlin, Germany\raisebox{11cm}[0cm][0cm]{\hspace*{5cm}\textbf{\ }}}\\[6mm]
\par\end{flushleft}

\begin{flushleft}
$^{3}$\parbox[t]{10cm}{Bernstein Center for Computational Neuroscience Berlin\\
Berlin, Germany\raisebox{11cm}[0cm][0cm]{\hspace*{5cm}\textbf{\ }}}\\[6mm]
\par\end{flushleft}

\vfill{}

\noindent{*} Correspondence:\hspace{1em}\parbox[t]{10cm}{
Thomas Pfeil\\
Universit{\"a}t Heidelberg\\
Kirchhoff-Institute for Physics\\
Im Neuenheimer Feld 227\\
69120 Heidelberg, Germany\\
tel: +49-6221-549813\\
\href{mailto:thomas.pfeil@kip.uni-heidelberg.de}{thomas.pfeil@kip.uni-heidelberg.de}}

\vfill{}

\noindent$^{\dagger}$ These authors contributed equally to this work. See acknowledgements for details.

\end{titlepage}
\begin{abstract}\thispagestyle{empty}\setcounter{page}{0}\pdfbookmark[1]{Abstract}{AbstractPage}

In this study, we present a highly configurable neuromorphic computing substrate and use it for emulating several types of neural networks.
%
At the heart of this system lies a mixed-signal chip, with analog implementations of neurons and synapses and digital transmission of action potentials.
%
Major advantages of this emulation device, which has been explicitly designed as a universal neural network emulator, are its inherent parallelism and high acceleration factor compared to conventional computers.
%
Its configurability allows the realization of almost arbitrary network topologies and the use of widely varied neuronal and synaptic parameters.
%
\verysmalltodominor{TP: I would prefer more glue here} Fixed-pattern noise inherent to analog circuitry is reduced by calibration routines.
%
\verysmalltodominor{TP: I would prefer more glue here} An integrated development environment allows neuroscientists to operate the device without any prior knowledge of neuromorphic circuit design.
%
As a showcase for the capabilities of the system, we describe the successful emulation of six different neural networks which cover a broad spectrum of both structure and functionality.

\bigskip{}

\noindent\textbf{Keywords:
accelerated neuromorphic hardware system,
universal computing substrate,
highly configurable,
mixed-signal VLSI,
spiking neural networks,
soft winner-take-all,
classifier,
cortical model.}
\verysmalltodominor{TP: discuss}

\end{abstract}
\clearpage{}
\section{Introduction}

By nature, computational neuroscience has a high demand for powerful and efficient devices for simulating neural network models.
In contrast to conventional general-purpose machines based on a von-Neumann architecture, neuromorphic systems are, in a rather broad definition, a class of devices which implement particular features of biological neural networks in their physical circuit layout \citep{Mead89_vlsibook, Indiveri09_119, Renaud10_905}.
In order to discern more easily between computational substrates, the term \emph{emulation} is generally used when referring to neural networks running on a neuromorphic back-end.

Several aspects motivate the neuromorphic approach.
The arguably most characteristic feature of neuromorphic devices is inherent parallelism enabled by the fact that individual neural network components (essentially neurons and synapses) are physically implemented \emph{in silico}.
Due to this parallelism, scaling of emulated network models does not imply slowdown, as is usually the case for conventional machines.
The hard upper bound in network size (given by the number of available components on the neuromorphic device) can be broken by scaling of the devices themselves, e.g., by wafer-scale integration \citep{Schemmel10_1947} or massively interconnected chips \citep{Merolla11_01}.
Emulations can be further accelerated by scaling down time constants compared to biology, which is enabled by deep submicron technology \citep{Schemmel06_1,Schemmel10_1947,Bruederle11_263}.
Unlike high-throughput computing with accelerated systems, real-time systems are often specialized for low power operation \citep[e.g.,][]{Indiveri06_211, Farquhar05_477}.

However, in contrast to the unlimited model flexibility offered by conventional simulation, the network topology and parameter space of neuromorphic systems are often dedicated for predefined applications and therefore rather restricted \citep[e.g.,][]{Serrano06_1217,Merolla06_4539,Akay07_book,Chicca07_981}.
Enlarging the configuration space always comes at the cost of hardware resources by occupying additional chip area.
Consequently, the maximum network size is reduced, or the configurability of one aspect is decreased by increasing the configurability of another.
Still, configurability costs can be counterbalanced by decreasing precision.
This could concern the size of integration time steps \citep{Imam12_25}, the granularity of particular parameters \citep{Pfeil12_90} or fixed-pattern noise affecting various network components.
At least the latter can be, to some extent, moderated through elaborate calibration methods \citep{Neftci10_262, Bruederle11_263, Goa12_2383}.

In this study, we present a user-friendly integrated development environment that can serve as a universal neuromorphic substrate for emulating different types of neural networks.
Apart from almost arbitrary network topologies, this system provides a vast configuration space for neuron and synapse parameters \citep{Schemmel06_1, Bruederle11_263}.
Reconfiguration is achieved on-chip and does not require additional support hardware.
While some models can easily be transferred from software simulations to the neuromorphic substrate, others need modifications.
These modifications take into account the limited hardware resources and compensate for fixed-pattern noise \citep{Kaplan09_1524, Bruederle09_17, Bruederle10_iscas, Bruederle11_263, Bill10_129}.
In the following, we show six more networks emulated on our hardware system, each requiring its own hardware configuration in terms of network topology and neuronal as well as synaptic parameters.
\section{The Neuromorphic System}
\label{sec:hw_system}

\begin{figure*}[tb]
\ifdraftmode
  \mbox{
  \hspace{-2.25cm}
\fi
\includegraphics[width=\fullfigwidth]{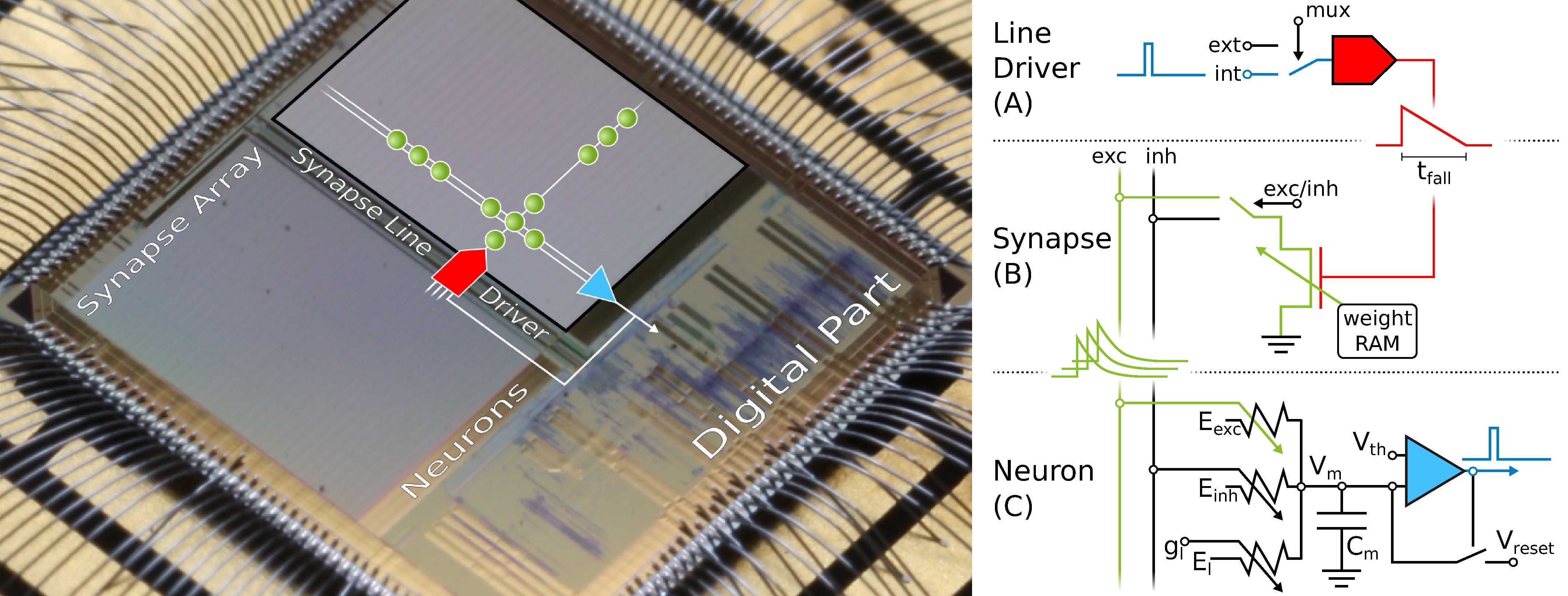}
\ifdraftmode
  }
\fi
\caption{\label{fig:spikey}
Microphotograph of the \spikey{} chip (fabricated in a \SI{180}{\nano\meter} CMOS process with die size $\SI{5}{} \times
\SI{5}{\milli\meter\squared}$).
Each of its 384 neurons can be arbitrarily connected to any other neuron.
In the following, we give a short overview of the technical implementation of neural networks on the \spikey{} chip.
\textbf{(A)}
Within the synapse array 256 synapse line drivers convert incoming digital spikes (blue) into a linear voltage ramp (red) with a falling slew rate $\tfall$.
For simplicity, the slew rate of the rising edge is not illustrated here, because it is chosen very small for all emulations in this study.
Each of these synapse line drivers are individually driven by either another on-chip neuron (int), e.g., the one depicted in (C),
or an external spike source (ext).
\textbf{(B)}
Within the synapse, depending on its individually configurable weight $\itsub{\weight}{i}$,
the linear voltage ramp (red) is then translated into a current pulse (green) with exponential decay.
These postsynaptic pulses are sent to the neuron via the excitatory (exc) and inhibitory (inh) input line,
shared by all synapses in that array column.
\textbf{(C)}
Upon reaching the neuron circuit, the total current on both input lines is converted into conductances, respectively.
If the membrane potential $\vmem$ crosses the firing threshold $\vth$, a digital pulse (blue) is generated,
which can be recorded and fed back into the synapse array.
After any spike, $\vmem$ is set to $\vreset$ for a refractory time period of $\tauref$.
Neuron and synapse line driver parameters can be configured as summarized in \prettyref{tab:paramlist}.
}
\end{figure*}

The central component of our neuromorphic hardware system is the neuromorphic microchip \spikey{}.
It contains analog very-large-scale integration (VLSI) circuits modeling the electrical behavior of neurons and synapses (\prettyref{fig:spikey}).
In such a \emph{physical model}, measurable quantities in the neuromorphic circuitry have corresponding biological equivalents.
For example, the membrane potential $\vmem$ of a neuron is modeled by the voltage over a capacitor $\cmem$ that, in turn, can be seen as a model of the capacitance of the cell membrane.
In contrast to numerical approaches, dynamics of physical quantities like $\vmem$ evolve continuously in time.
We designed our hardware systems to have time constants approximately $10^4$ times faster than their biological counterparts allowing for high-throughput computing.
This is achieved by reducing the size and hence the time constant of electrical components, which also allows for more neurons and synapses on a single chip.
To avoid confusion between hardware and biological domains of time, voltages and currents, all parameters are specified in biological domains throughout this study.

\subsection{The Neuromorphic Chip}
\label{sec:hw_neuron}

On \spikey{} (\prettyref{fig:spikey}), a VLSI version of the standard leaky integrate-and-fire (LIF) neuron model with conductance-based synapses is implemented \citep{Dayan01}:
\begin{equation}
\label{eq:lif}
\cmem \frac{\mathrm{d}\vmem}{\mathrm{d}t} = - \gl (\vmem - \el) - \sum_i \itsub{g}{i} (\vmem - \itsub{E}{i})
\end{equation}
For its hardware implementation see \prettyref{fig:spikey}, \citet{Schemmel06_1} and \citet{Indiveri11_73}.

Synaptic conductances $\itsub{g}{i}$ (with the index $i$ running over all synapses) drive the membrane potential $\vmem$ towards the reversal potential $\itsub{E}{i}$, with $\itsub{E}{i} \in \{{\eexc,\einh}\}$.
The time course of the synaptic activation is modeled by
\begin{equation}
\label{eq:tot_gsyn}
\itsub{g}{i}(t) = \itsub{p}{i}(t) \cdot \itsub{\weight}{i} \cdot \gimax
\end{equation}
where $\gimax$ are the maximum conductances and $\itsub{\weight}{i}$ the weights for each synapse, respectively.
The time course $\itsub{p}{i}(t)$ of synaptic conductances is a linear transformation of the current pulses shown in \prettyref{fig:spikey} (green), and hence an exponentially decaying function of time.
The generation of conductances at the neuron side is described in detail by \citet{Indiveri11_73}, postsynaptic potentials are measured by \citet{Schemmel07_iscas}.

The implementation of spike-timing dependent plasticity \citep[STDP;][]{Bi98,Song00} modulating $\itsub{\weight}{i}$ over time is described in \citet{Schemmel06_1} and \citet{Pfeil12_90}.
Correlation measurement between pre- and post-synaptic action potentials is carried out in each synapse,
and the 4-bit weight is updated by an on-chip controller located in the digital part of the \spikey{} chip.
However, STDP will not be further discussed in this study.

Short-term plasticity (STP) modulates $\gimax$ \citep{Schemmel07_iscas} similar to the model by \citet{Tsodyks97} and \citet{Markram98}.
On hardware, STP can be configured individually for each synapse line driver that corresponds to an axonal connection in biological terms.
It can either be facilitating or depressing.

The propagation of spikes within the \spikey{} chip is illustrated in \prettyref{fig:spikey} and described in detail by \citet{Schemmel06_1}.
\emph{Spikes} enter the chip as time-stamped events using standard digital signaling techniques that facilitate long-range communication, e.g., to the host computer or other chips.
Such digital packets are processed in discrete time in the digital part of the chip, where they are transformed into digital \emph{pulses} entering the synapse line driver (blue in \prettyref{fig:spikey}A).
These pulses propagate in continuous time between on-chip neurons, and are optionally transformed back into digital spike packets for off-chip communication.

\subsection{System Environment}
\label{sec:hw_environment}

\begin{figure}[t] 
	\centering
	\includegraphics[width=\columnfigwidth]{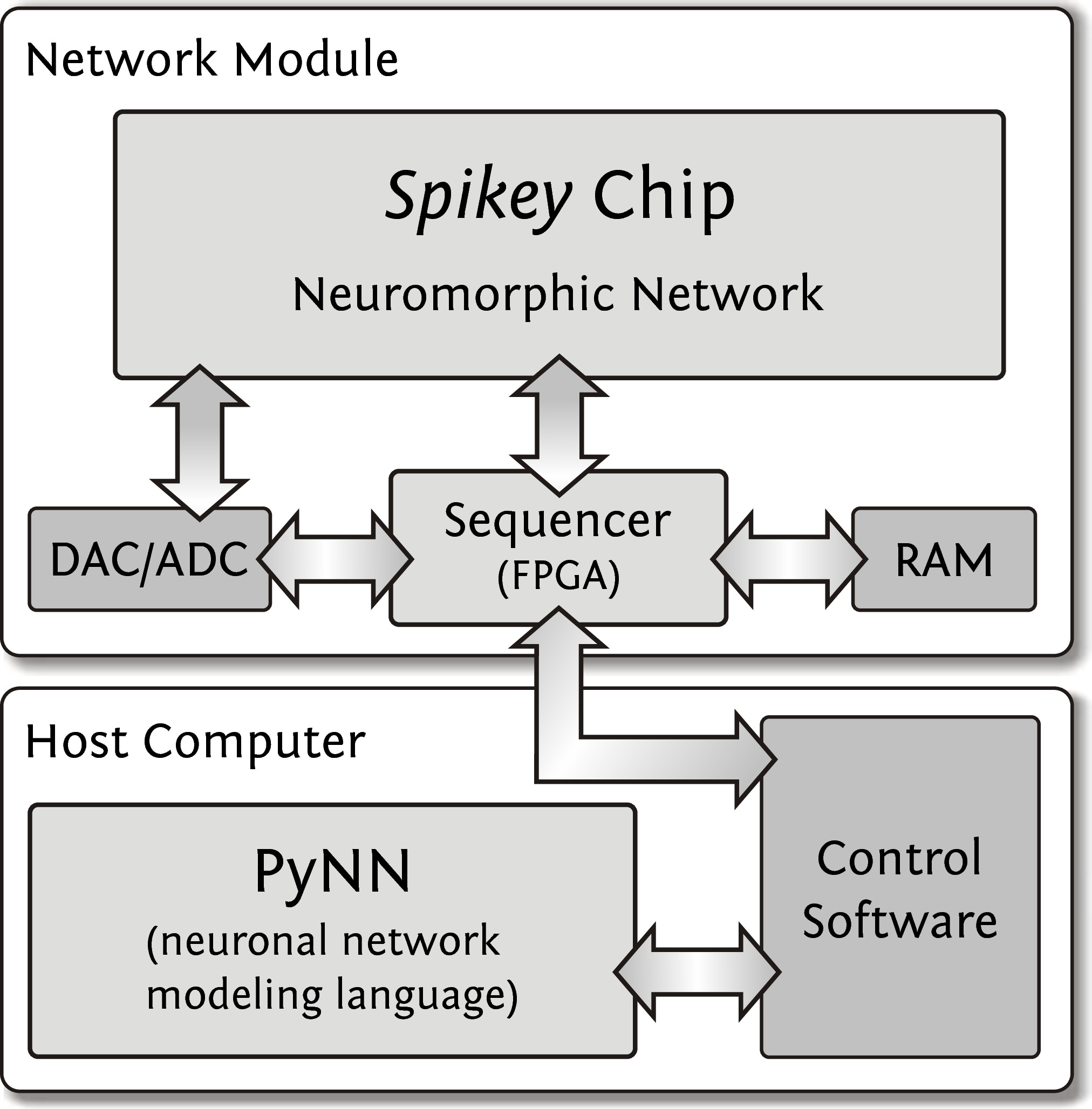}
	\caption{\label{fig:hw_setup}
	Integrated development environment:
	User access to the \spikey{} chip is provided using the \swpackage{PyNN} neural network modeling language.
	The control software controls and interacts with the network module which is operating the \spikey{} chip.
  The RAM size (\SI{512}{\mega\byte}) limits the total number of spikes for stimulus and spike recordings to approx.\ $2 \cdot 10^8$ spikes.
  The required data for a full configuration of the \spikey{} chip has a size of approx.\ $\SI{100}{\kilo\byte}$.
	}
\end{figure}

\verysmalltodominor{TP: thought about swapping the sections Config with System Environment, here I can put some glue that system environment fundamental for configuration, already stated somewhere else?}
The \spikey{} chip is mounted on a network module described and schematized in \citet{Fieres04_bics} and \prettyref{fig:hw_setup}, respectively.
Digital spike and configuration data is transferred via direct connections between a field-programmable gate array (FPGA) and the \spikey{} chip.
Onboard digital-to-analog converter (DAC) and analog-to-digital converter (ADC) components supply external parameter voltages to the \spikey{} chip and digitize selected voltages generated by the chip for calibration purposes.
Furthermore, up to eight selected membrane voltages can be recorded in parallel by an oscilloscope.
Because communication between a host computer and the FPGA has a limited bandwidth that does not satisfy real-time operation requirements of the \spikey{} chip, experiment execution is controlled by the FPGA while operating the \spikey{} chip in continuous time.
To this end, all experiment data is stored in the local random access memory (RAM) of the network module.
Once the experiment data is transferred to the local RAM, emulations run with an acceleration factor of $10^4$ compared to biological real-time.
This acceleration factor applies to all emulations shown in this study, independent of the size of networks.

Execution of an experiment is split up into three steps (\prettyref{fig:hw_setup}).
First, the \emph{control software} within the memory of the host computer generates configuration data (\prettyref{tab:paramlist},
e.g., synaptic weights, network connectivity, etc.), as well as input stimuli to the network.
All data is stored as a sequence of commands and is transferred to the memory on the network module.
In the second step, a playback sequencer in the FPGA logic interprets this data and sends it to the \spikey{} chip, as well as triggers the emulation.
Data produced by the chip, e.g., neuronal activity in terms of spike times, is recorded in parallel.
In the third and final step, this recorded data stored in the memory on the network module is retrieved and transmitted to the host computer, where they are processed by the control software.

Having a control software that abstracts hardware greatly simplifies modeling on the neuromorphic hardware system.
However, modelers are already struggling with multiple incompatible interfaces to software simulators.
That is why our neuromorphic hardware system supports \swpackage{PyNN}, a widely used application programming interface (API) that strives for a coherent user interface,
allowing portability of neural network models between different software simulation frameworks (e.g., \swpackage{NEST} or \swpackage{NEURON}) and hardware systems (e.g., the \spikey{} system).
For details see
\citet{Gewaltig_07_11204,Eppler09_12} for \swpackage{NEST},
\citet{Carnevale06,Hines09} for \swpackage{NEURON},
\citet{Bruederle11_263,Bruederle09_17} for the \spikey{} chip, and
\citet{Davison09,Davison10_pynn} for \swpackage{PyNN}, respectively.

\subsection{Configurability}
\label{sec:hw_configurability}

\newlength{\multirowwidth}
\begin{table*}[tbp]
	\newlength{\saveTabColSep}
	\setlength{\saveTabColSep}{\tabcolsep}
	\setlength{\tabcolsep}{0pt}
\centering
\mbox{
\ifdraftmode
  \hspace{-2.75cm}
  \footnotesize
  \setlength{\multirowwidth}{2cm}
  \begin{tabular}{L{\multirowwidth} C{2cm} C{1cm} L{14cm}}
\else
  \small
  \hspace{-0.45cm}
  \setlength{\multirowwidth}{2cm}
  \begin{tabular}{L{\multirowwidth} C{2cm} C{1cm} L{13cm}} 
\fi
\toprule
Scope                                                    & Name                   & Type               & Description \\
\midrule\midrule
\multirow{8}{\multirowwidth}{Neuron circuits~(A)}        & n/a                    & $\paramtype{i}{n}$ & Two digital configuration bits activating the neuron and readout of its membrane voltage\\
                                                         & $\gl$                  & $\paramtype{i}{n}$ & Bias current for neuron leakage circuit\\ 
                                                         & $\tauref$              & $\paramtype{i}{n}$ & Bias current controlling neuron refractory time \\
                                                         & $\el$                  & $\paramtype{s}{n}$ & Leakage reversal potential\\
                                                         & $\einh$                & $\paramtype{s}{n}$ & Inhibitory reversal potential\\
                                                         & $\eexc$                & $\paramtype{s}{n}$ & Excitatory reversal potential\\
                                                         & $\vth$                 & $\paramtype{s}{n}$ & Firing threshold voltage\\
                                                         & $\vreset$              & $\paramtype{s}{n}$ & Reset potential\\
\midrule
\multirow{4}{\multirowwidth}{Synapse line drivers~(B)}   & n/a                    & $\paramtype{i}{l}$ & Two digital configuration bits selecting input of line driver\\
                                                         & n/a                    & $\paramtype{i}{l}$ & Two digital configuration bits setting line excitatory or inhibitory\\
                                                         & $\trise,\tfall$        & $\paramtype{i}{l}$ & Two bias currents for rising and falling slew rate of presynaptic voltage ramp\\ 
                                                         & $\gimax$               & $\paramtype{i}{l}$ & Bias current controlling maximum voltage of presynaptic voltage ramp\\ 
\midrule
Synapses~(B)                                             & $\weight$              & $\paramtype{i}{s}$ & 4-bit weight of each individual synapse\\
\midrule
\multirow{6}{\multirowwidth}{STP related~(C)}            & n/a                    & $\paramtype{i}{l}$ & Two digital configuration bits selecting short-term depression or facilitation\\
                                                         & $\mrmsub{U}{SE}$       & $\paramtype{i}{l}$ & Two digital configuration bits tuning synaptic efficacy for STP\\
                                                         & n/a                    & $\paramtype{s}{l}$ & Bias voltage controlling spike driver pulse length\\ 
                                                         & $\taurec$, $\taufacil$ & $\paramtype{s}{l}$ & Voltage controlling STP time constant\\ 
                                                         & I                      & $\paramtype{s}{l}$ & Short-term facilitation reference voltage\\ 
                                                         & R                      & $\paramtype{s}{l}$ & Short-term capacitor high potential\\ 
\midrule
\multirow{4}{\multirowwidth}{STDP related~(D)}           & n/a                    & $\paramtype{i}{l}$ & Bias current controlling delay for presynaptic correlation pulse (for calibration purposes)\\ 
                                                         & $\aplusminus$          & $\paramtype{s}{l}$ & Two voltages dimensioning charge accumulation per (anti-)causal correlation measurement\\ 
                                                         & n/a                    & $\paramtype{s}{l}$ & Two threshold voltages for detection of relevant (anti-)causal correlation\\ 
                                                         & $\taustdp$             & $\paramtype{g}{}$  & Voltage controlling STDP time constants\\ 
\bottomrule
\end{tabular}
}
\caption{\label{tab:paramlist} List of analog current and voltage parameters as well as digital configuration bits. Each with corresponding model parameter names, excluding technical parameters that are only relevant for correctly biasing analog support circuitry or controlling digital chip functionality.
Electronic parameters that have no direct translation to model parameters are denoted \emph{n/a}.
The membrane capacitance is fixed and identical for all neuron circuits ($\cmem=\SI{0.2}{\nano\farad}$ in biological value domain).
Parameter types:
(i) controllable for each corresponding circuit: 192 for neuron circuits (denoted with subscript n), 256 for synapse line drivers (denoted with subscript l), 49152 for synapses (denoted with subscript s),
(s) two values, shared for all even/odd neuron circuits or synapse line drivers, respectively,
(g) global, one value for all corresponding circuits on the chip.
All numbers refer to circuits associated to one synapse array and are doubled for the whole chip.
For technical reasons, the current revision of the chip only allows usage of one synapse array of the chip.
Therefore, all experiments presented in this paper are limited to a maximum of 192 neurons.
For parameters denoted by (A) see \prettyref{eq:lif} and \citet{Schemmel06_1}, for (B) see \prettyref{fig:spikey}, \prettyref{eq:tot_gsyn} and \citet{Dayan01}, for (C) see \citet{Schemmel07_iscas} and for (D) see \citet{Schemmel06_1} and \citet{Pfeil12_90}.
\smalltodominor[inline]{JS -> TP: replace causal with pre-post and anti-causal with post-pre; TP: more characters resulting in line break :-(}
\smalltodominor[inline]{JS -> TP: vclra/c and vcthigh/low could be put together}
}
\setlength{\tabcolsep}{\saveTabColSep}
\end{table*}

In order to facilitate the emulation of network models inspired by biological neural structures, it is essential to support the implementation of different (cortical) neuron types.
From a mathematical perspective, this can be achieved by varying the appropriate parameters of the implemented neuron model (\prettyref{eq:lif}).

To this end, the \spikey{} chip provides 2969 different analog parameters (\prettyref{tab:paramlist}) stored on current memory cells that are continuously refreshed from a digital on-chip memory.
Most of these cells deliver individual parameters for each neuron (or synapse line driver), e.g., leakage conductances $\gl$.
Due to the size of the current-voltage conversion circuitry it was not possible to provide individual voltage parameters, such as, e.g., $\el$, $\eexc$ and $\einh$, for each neuron.
As a consequence, groups of 96 neurons share most of these voltage parameters.
Parameters that can not be controlled individually are delivered by global current memory cells.

In addition to the possibility of controlling analog parameters, the \spikey{} chip also offers an almost arbitrary configurability of the network topology.
As illustrated in \prettyref{fig:spikey}, the fully configurable \emph{synapse array} allows connections from synapse line drivers (located alongside the array) to arbitrary neurons (located below the array) via synapses whose weights can be set individually with a 4-bit resolution.
This limits the maximum fan-in to 256 synapses per neuron, which can be composed of up to 192 synapses from on-chip neurons, and up to 256 synapses from external spike sources.
Because the total number of neurons exceeds the number of inputs per neuron, an all-to-all connectivity is not possible.
For all networks presented in this study, the connection density is much lower than realizable on the chip, which supports the chosen trade-off between inputs per neuron and total neuron count.

\subsection{Calibration}
\label{sec:hw_calibration}

Device mismatch that arises from hardware production variability causes fixed-pattern noise, which causes parameters to vary from neuron to neuron as well as from synapse to synapse.
Electronic noise (including thermal noise) also affects dynamic variables, as, e.g., the membrane potential $\vmem$.
Consequently, experiments will exhibit some amount of both neuron-to-neuron and trial-to-trial variability given the same input stimulus.
It is, however, important to note that these types of variations are not unlike the neuron diversity and response stochasticity found in biology \citep{Gupta00, Maass02_2531, Marder06_563, rolls10_noisybook}.

To facilitate modeling and provide repeatability of experiments on arbitrary \spikey{} chips, it is essential to minimize these effects by calibration routines.
Many calibrations have directly corresponding biological model parameters, e.g., membrane time constants (described in the following), firing thresholds, synaptic efficacies or PSP shapes.
Others have no equivalents, like compensations for shared parameters or workarounds of defects \citep[e.g.,][]{Kaplan09_1524,Bill10_129,Pfeil12_90}.
In general, calibration results are used to improve the mapping between biological input parameters and the corresponding target hardware voltages and currents, as well as to determine the dynamic range of all model parameters \citep[e.g.,][]{Bruederle09_17}.

\begin{figure}[t] 
	\centering
	\includegraphics[width=\columnfigwidth,clip]{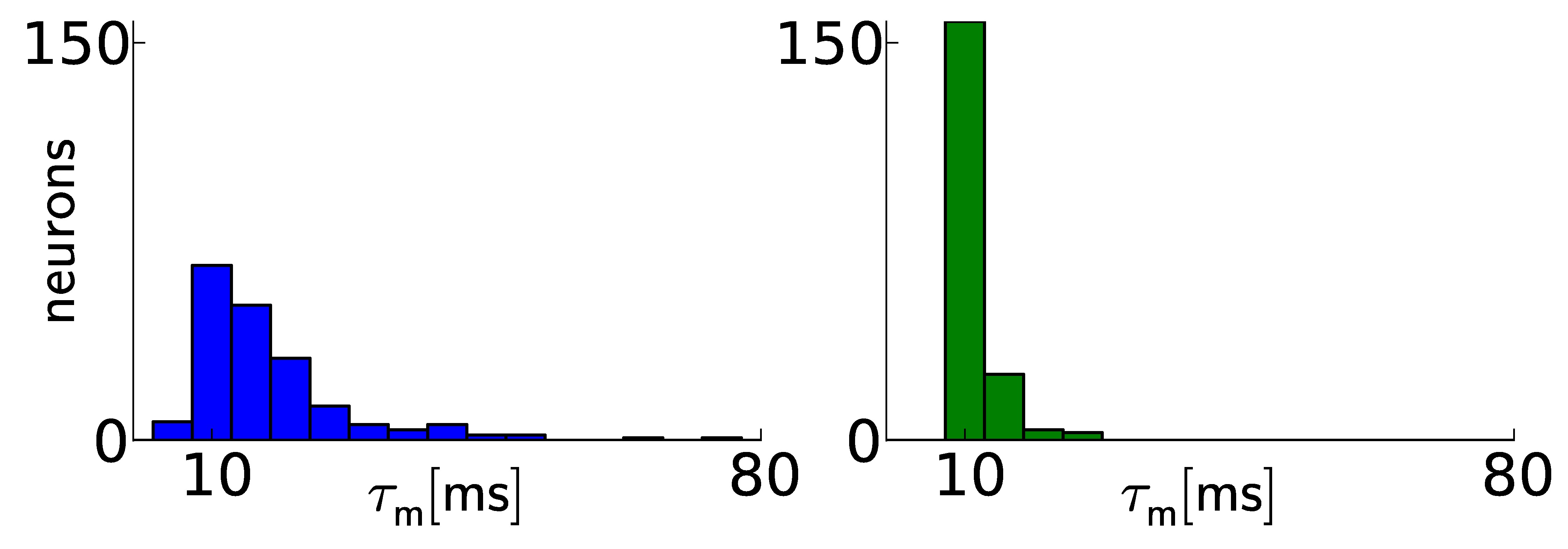}
	\caption{\label{fig:taumem_hist}
	Calibration results for membrane time constants: Before calibration (left), the distribution of $\taum$ values has a median of $\widetilde{\taum} = \SI{15.1}{\milli\second}$ with 20th and 80th percentiles of $\taum^{20} = \SI{10.3}{\milli\second}$ and $\taum^{80} = \SI{22.1}{\milli\second}$, respectively.
	After calibration (right), the distribution median lies closer to the target value and narrows significantly: $\widetilde{\taum} = \SI{11.2}{\milli\second}$ with $\taum^{20} = \SI{10.6}{\milli\second}$ and $\taum^{80} = \SI{12.0}{\milli\second}$.
  Two neurons were discarded, because the automated calibration algorithm did not converge.
	}
\end{figure}

While the calibration of most parameters is rather technical, but straightforward (e.g., all neuron voltage parameters), some require more elaborate techniques.
These include the calibration of $\taum$, STP as well as synapse line drivers, as we describe later for individual network models.
The membrane time constant $\taum = \cmem / \gl$ differs from neuron to neuron mostly due to variations in the leakage conductance $\gl$.
However, $\gl$ is independently adjustable for every neuron.
Because this conductance is not directly measurable, an indirect calibration method is employed.
To this end, the threshold potential is set below the resting potential.
Following each spike, the membrane potential is clamped to $\vreset$ for an absolute refractory time  $\tauref$, after which it evolves exponentially towards the resting potential $\el$ until the threshold voltage triggers a spike and the next cycle begins.
If the threshold voltage is set to $\vth = \el - 1/e \cdot (\el - \vreset)$, the spike frequency equals $1/(\taum + \tauref)$, thereby allowing an indirect measurement and calibration of $\gl$ and therefore $\taum$.
For a given $\taum$ and $\tauref = const$, $\vth$ can be calculated.
An iterative method is applied to find the best-matching $\vth$,
because the exact hardware values for $\el$, $\vreset$ and $\vth$ are only known after the measurement.
The effect of calibration on a typical chip can best be exemplified for a typical target value of $\taum = \SI{10}{\milli\second}$.
\prettyref{fig:taumem_hist} depicts the distribution of $\taum$ of a typical chip before and after calibration.

The STP hardware parameters have no direct translation to model equivalents.
In fact, the implemented transconductance amplifier tends to easily saturate within the available hardware parameter ranges.
These non-linear saturation effects can be hard to handle in an automated fashion on an individual circuit basis.
Consequently, the translation of these parameters is based on STP courses averaged over several circuits.
\section{Hardware Emulation of Neural Networks}

In the following, we present six neural network models that have been emulated on the \spikey{} chip.
Most of the emulation results are compared to those obtained by software simulations in order to verify the network functionality and performance.
For all these simulations the tool \swpackage{NEST} \citep{Gewaltig_07_11204} or \swpackage{NEURON} \citep{Carnevale06} is used.

\subsection{Synfire Chain with Feedforward Inhibition}
\label{sec:sfc}

\ifdraftmode
\else
\begin{figure*}[tb]
\ifdraftmode
  \mbox{
  \hspace{-2.25cm}
\fi
\includegraphics[width=\fullfigwidth]{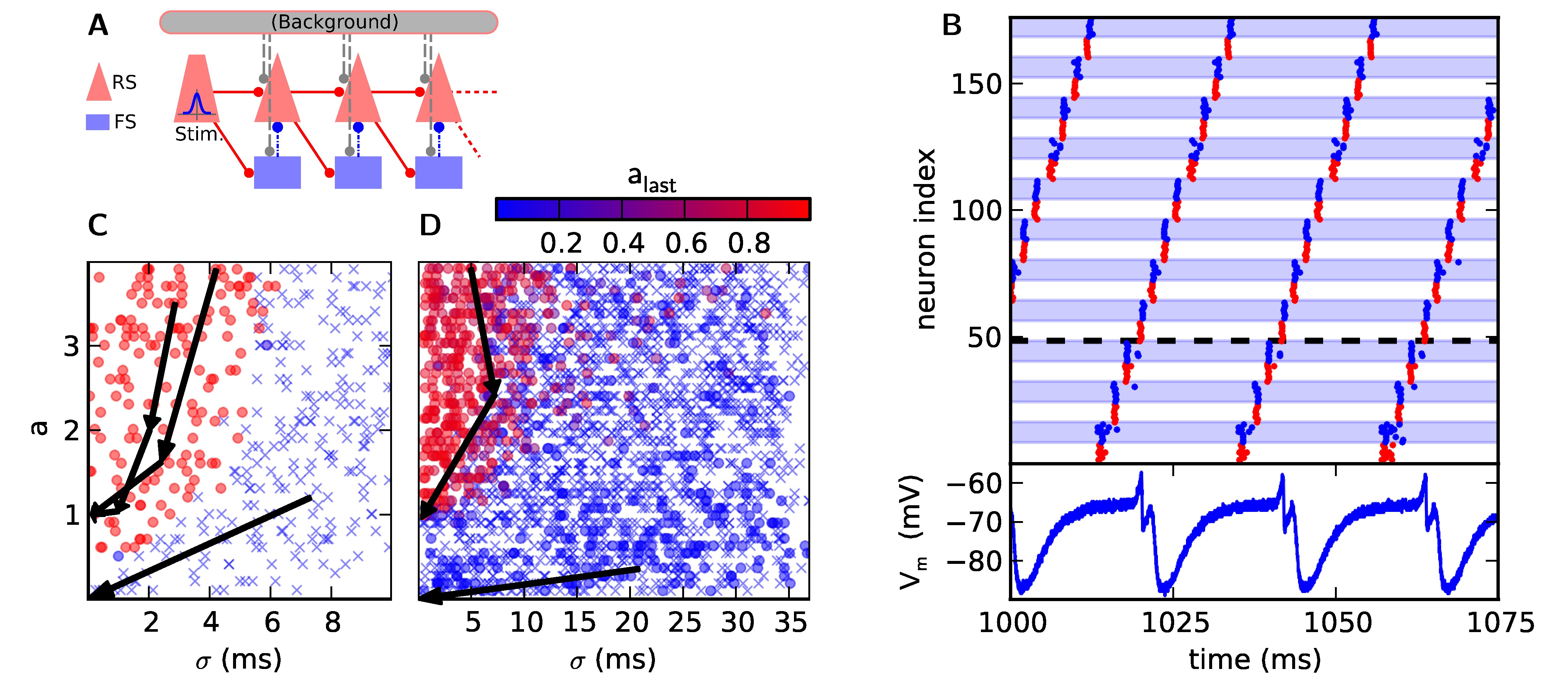}
\ifdraftmode
  }
\fi
\caption{
\textbf{(A)}
Synfire chain with feedforward
inhibition.
The background is only utilized in the original model,
where it is implemented as random Gaussian current.
For the presented hardware implementation it has been
omitted due to network size constraints.
As compensation for missing background stimuli, the resting potential was increased to ensure a comparable
excitability of the neurons.
\textbf{(B)}
Hardware emulation.
Top:
Raster plot of pulse packet propagation
\SI{1000}{\milli\second} after initial stimulus.
Spikes from RS groups are shown in red and spikes from FS groups are
denoted by blue color and background.
Bottom:
Membrane potential of the first neuron in the fourth RS group,
which is denoted by a dashed horizontal line.
The cycle duration is approximately \SI{20}{\milli\second}.
\textbf{(C)}
State space generated with software simulations of the original model.
The position of each marker indicates the $(\sigma, a)$ parameters of
the stimulus while the color encodes the activity in the RS population
of the third synfire group.
Lack of activity is indicated with a cross.
The evolution of the pulse packet parameters is shown for three
selected cases by a series of arrows.
Activity either stably propagates with fixed point $(\sigma, a) = (\SI{0.1}{\milli\second}, 1)$
or extinguishes with fixed point $(\sigma, a) = (\SI{0}{\milli\second}, 0)$.
\textbf{(D)}
Same as (C), but emulated on the FACETS chip-based system.
The activity in the last group is located either near $(\sigma, a) = (\SI{0}{\milli\second}, 0)$
or $(\SI{0.3}{\milli\second}, 1)$.
The difference to software simulations is explained in \prettyref{sec:synfire_hardware}.
\label{fig:synfire_chain}
\smalltodominor[inline]{TP: minor minor beauty changes: remove border in A; unify with BRN figure?} 
}
\end{figure*}
\fi

Architectures with a feedforward connectivity have been employed
extensively as computational components and as models for the study of
neuronal dynamics.
Synfire chains are feedforward networks consisting of several neuron
groups where each neuron in a group projects to neurons in the
succeeding group.

They have been originally proposed to account for the presence of
behaviorally-related, highly precise firing patterns \citep{Baker01b, Prut98}.
Further properties of such structures have been studied extensively,
including activity transport \citep{Aertsen96b, Diesmann99, Litvak03},
external control of information flow
\citep{Kremkow10_15760}, computational capabilities
\citep{Abeles04_179, Vogels05b,
Schrader-2010_online}, complex dynamic behavior
\citep{Yazdanbakhsh02_367} and their embedding into surrounding networks
\citep{Aviel03a, Tetzlaff05a,
Schrader08}.
\citet{Kremkow-2010} have shown that feedforward
inhibition can increase the selectivity to the initial stimulus and
that the local delay of inhibition can modify this selectivity.

\subsubsection{\network{}}
\label{sec:synfire_network}

The presented network model is an adaptation of the
\emph{feedforward network} described in \citet{Kremkow-2010}.

The network consists of several neuron groups, each comprising
$\mrmsub{n}{RS}=100$ excitatory regular spiking (RS) and
$\mrmsub{n}{FS}=25$ inhibitory fast spiking (FS) cells.
All neurons are modeled as LIF neurons with
exponentially decaying synaptic conductance courses.
According to \citet{Kremkow-2010} all neurons have identical parameters.

As shown in \prettyref{fig:synfire_chain}A, RS neurons project
to both RS and FS populations in the subsequent group while the FS population
projects to the RS population in its local group.
Each neuron receives a fixed number of randomly chosen inputs from
each presynaptic population.
The first group is stimulated by a population of $\mrmsub{n}{RS}$
external spike sources with identical connection probabilities as used for RS groups
within the chain.

Two different criteria are employed to assess the functionality of the
emulated synfire chain.
The first, straightforward benchmark is the stability of signal
propagation.
An initial synchronous stimulus is expected to cause a stable
propagation of activity, with each neuron in an RS population
spiking exactly once.
Deviations from the original network parameters can cause the activity to grow rapidly,
i.e., each population emits more spikes than its predecessor,
or stall pulse propagation.

The second, broader characterization follows
\citet{Kremkow-2010}, who has analyzed the response of the network to various
stimuli.
The stimulus is parametrized by the variables $a$ and $\sigma$.
For each neuron in the stimulus population $a$ spike times are
generated by sampling them from a Gaussian distribution with common
mean and standard deviation.
$\sigma$ is defined as the standard deviation of the spike times of
all source neurons.
Spiking activity that is evoked in the subsequent RS populations is
characterized analogously by measuring $a$ and $\sigma$.

\prettyref{fig:synfire_chain}C shows the result of a
software simulation of the original network.
%
%
The filter properties of the network are reflected by a separatrix dividing the state space shown in \prettyref{fig:synfire_chain}C and D into two areas, each with a different fixed point.
First, the basin of attraction (dominated by red circles in \prettyref{fig:synfire_chain}C) from which stable propagation can be evoked and second, the remaining region (dominated by crosses in \prettyref{fig:synfire_chain}C) where any initial activity becomes extinguished.
This separatrix determines which types of initial input
lead to a stable signal propagation.

\subsubsection{\hardware{}}
\label{sec:synfire_hardware}

\ifdraftmode
\begin{figure*}[tb]
\ifdraftmode
  \mbox{
  \hspace{-2.25cm}
\fi
\includegraphics[width=\fullfigwidth]{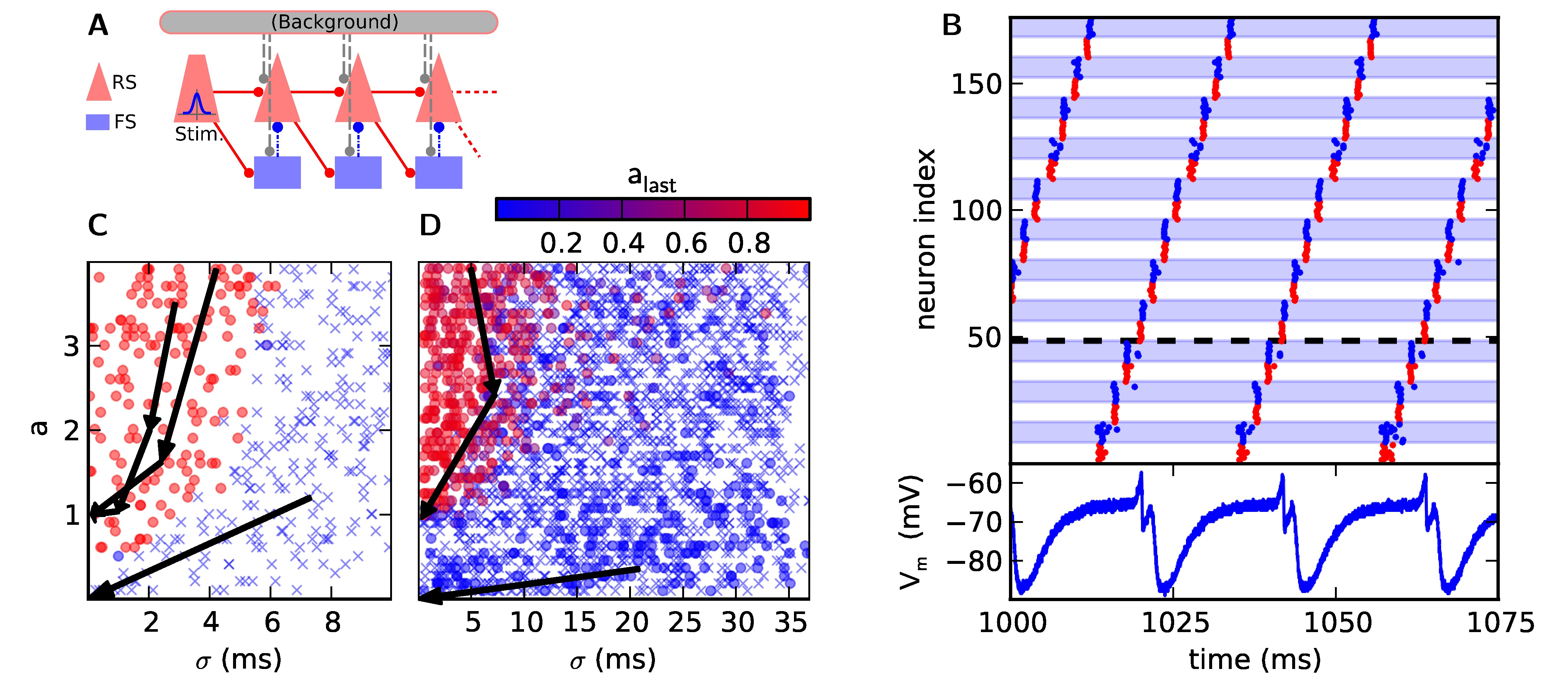}
\ifdraftmode
  }
\fi
\caption{
\textbf{(A)}
Synfire chain with feedforward
inhibition.
The background is only utilized in the original model,
where it is implemented as random Gaussian current.
For the presented hardware implementation it has been
omitted due to network size constraints.
As compensation for missing background stimuli, the resting potential was increased to ensure a comparable
excitability of the neurons.
\textbf{(B)}
Hardware emulation.
Top:
Raster plot of pulse packet propagation
\SI{1000}{\milli\second} after initial stimulus.
Spikes from RS groups are shown in red and spikes from FS groups are
denoted by blue color and background.
Bottom:
Membrane potential of the first neuron in the fourth RS group,
which is denoted by a dashed horizontal line.
The cycle duration is approximately \SI{20}{\milli\second}.
\textbf{(C)}
State space generated with software simulations of the original model.
The position of each marker indicates the $(\sigma, a)$ parameters of
the stimulus while the color encodes the activity in the RS population
of the third synfire group.
Lack of activity is indicated with a cross.
The evolution of the pulse packet parameters is shown for three
selected cases by a series of arrows.
Activity either stably propagates with fixed point $(\sigma, a) = (\SI{0.1}{\milli\second}, 1)$
or extinguishes with fixed point $(\sigma, a) = (\SI{0}{\milli\second}, 0)$.
\textbf{(D)}
Same as (C), but emulated on the FACETS chip-based system.
The activity in the last group is located either near $(\sigma, a) = (\SI{0}{\milli\second}, 0)$
or $(\SI{0.3}{\milli\second}, 1)$.
The difference to software simulations is explained in \prettyref{sec:synfire_hardware}.
\label{fig:synfire_chain}
\smalltodominor[inline]{TP: minor minor beauty changes: remove border in A; unify with BRN figure?} 
}
\end{figure*}
\fi

The original network model could not be mapped directly to the
\spikey{} chip because it requires 125 neurons per group, while on the
chip only 192 neuron circuits are available.
Further constraints were caused by the fixed synaptic delays, which are
determined by the speed of signal propagation on the chip.
The magnitude of the delay is approximately \SI{1}{\milli\second} in biological time.

By simple modifications of the network, we were able to qualitatively
reproduce both benchmarks defined in \prettyref{sec:synfire_network}.
%
Two different network configurations were used, each adjusted to the requirements of one benchmark.
In the following, we describe these differences, as well as the results for each benchmark.

To demonstrate a stable propagation of pulses, a large number of consecutive group
activations was needed.
The chain was configured as a loop by connecting the last group to the
first, allowing the observation of more pulse packet propagations than there are
groups in the network.
%
%

The time between two passes of the pulse packet at the same synfire
group needs to be maximized to allow the neurons to recover
(see voltage trace in \prettyref{fig:synfire_chain}B).
This is accomplished by increasing the group count and consequently reducing the group size.
As too small populations cause an unreliable signal propagation, which
is mainly caused by inhomogeneities in the neuron behavior,
$\mrmsub{n}{RS}=\mrmsub{n}{FS}=8$ was chosen as a satisfactory
trade-off between propagation stability and group size.
Likewise, the proportion of FS neurons in a group was increased to
maintain a reliable inhibition.
To further improve propagation properties, the membrane time constant
was lowered for all neurons by raising $\gl$ to its maximum value.
The strength of inhibition was increased by
setting the inhibitory synaptic weight to its maximum value
and lowering the inhibitory reversal potential to its minimum value.
Finally, the synaptic weights $\mrmsub{RS}{i} \to \mrmsub{RS}{i + 1}$
and $\mrmsub{RS}{i} \to \mrmsub{FS}{i + 1}$ were adjusted.
%
With these improvements we could observe persisting synfire propagation on the oscilloscope \SI{2}{\hour} wall-clock
time after stimulation.
This corresponds to more than 2 years in biological real-time.

The second network demonstrates the filtering properties of a hardware-emulated
synfire chain with feedforward inhibition.
This use case required larger synfire groups than in the first case as
otherwise, the total excitatory conductance caused by a pulse packet
with large $\sigma$ was usually not smooth enough due to the
low number of spikes.
Thus, three groups were placed on a single chip with
$\mrmsub{n}{RS}=45$ and $\mrmsub{n}{FS}=18$.
The resulting evolution of pulse packets is shown in
\prettyref{fig:synfire_chain}D.
After passing three groups, most runs resulted in either very low
activity in the last group or were located near the point
$(\SI{0.3}{\milli\second}, 1)$, as illustrated in \prettyref{fig:synfire_chain}D.

Emulations on hardware differ from software simulations in two important points:
First, the separation in the parameter space of the initial stimulus
is not as sharply bounded, which is demonstrated by the fact that
occasionally, significant activity in the last group can be evoked by
stimuli with large $\sigma$ and large $a$, as seen in
\prettyref{fig:synfire_chain}D.
This is a combined effect due to the reduced population sizes and the
fixed pattern noise in the neuronal and synaptic circuits.
Second, a stimulus with a small $a$ can evoke weak activity in the
last group, which is attributed to a differing balance between
excitation and inhibition.
In hardware, a weak stimulus causes both, the RS and FS populations to
response weakly which leads to a weak inhibition of the RS population,
allowing the pulse to reach the last synfire group.
Hence, the pulse fades slowly instead of being extinguished completely.
In the original model, the FS population is more responsive and
prevents the propagation more efficiently.
%


%
Nevertheless, the filtering properties of the network are apparent.
The quality of the filter
could be improved by employing the original group size,
which would require using a large-scale neuromorphic device
\citep[see, e.g.,][]{Schemmel10_1947}.

Our hardware implementation of the synfire chain model demonstrates the
possibility to run extremely long lasting experiments due to the high acceleration factor of the hardware system.
Because the synfire chain model itself does not require sustained external
stimulus, it could be employed as an autonomous source of periodic input to other
experiments.
\subsection{Balanced Random Network}
\label{sec:brn}

\ifdraftmode
\else
\begin{figure*}[tb]
\ifdraftmode
  \mbox{
  \hspace{-2.25cm}
\fi
\includegraphics[width=\fullfigwidth]{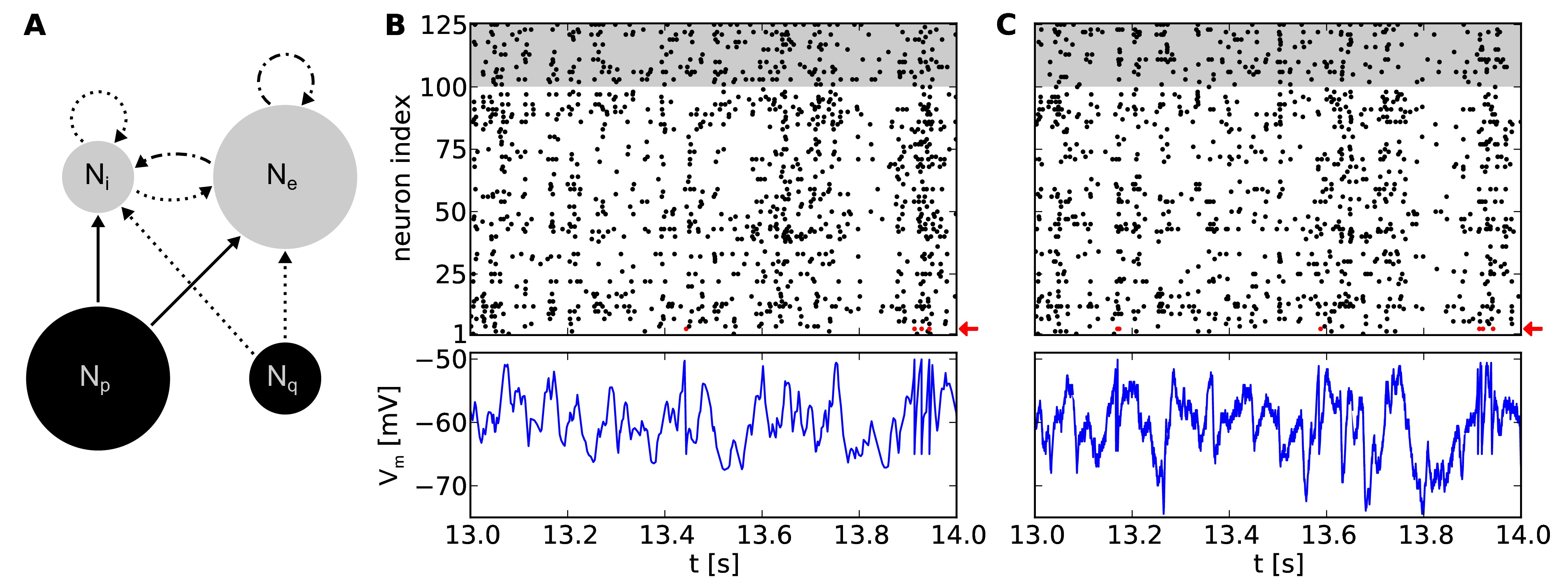}
\ifdraftmode
  }
\fi
\caption{
\textbf{(A)}
Network topology of a balanced random network.
Populations consisting of $\mrmsub{N}{e}=100$ excitatory and $\mrmsub{N}{i}=25$ inhibitory neurons (gray circles), respectively, are stimulated by populations of Poisson sources (black circles).
We use $\mrmsub{N}{p}=100$ independent sources for excitation and $\mrmsub{N}{q}=25$ for inhibition.
Arrows denote projections between these populations with connection probabilities $p=0.1$, with solid lines for excitatory and dotted lines for inhibitory connections.
Dot and dash lines are indicating excitatory projections with short-term depression.
\textbf{(B)}
Top:
Raster plot of a software simulation.
Populations of excitatory and inhibitory neurons are depicted with white and gray background, respectively.
Note that for clarity only the time interval [\SI{13}{\second},~\SI{14}{\second}] of a \SI{20}{\second} emulation is shown.
For the full \SI{20}{\second} emulation, we have measured $\cv=0.96\pm0.09$ (mean over all neurons) and $\cc=0.010\pm0.017$ (mean over 1000 random chosen pairs of neurons), respectively.
Bottom:
Recorded membrane potential of an arbitrary excitatory neuron (neuron index 3, highlighted with a red arrow in the above raster plot).
\textbf{(C)}
Same network topology and stimulus as in (B), but emulated on the \spikey{} chip, resulting in $\cv=1.02\pm0.16$ and $\cc=0.014\pm0.019$.
Note that the membrane recordings are calibrated such that the threshold and reset potential match those of the software counterpart.
\label{fig:brn}}
\end{figure*}
\fi

\smalltodominor[inline]{MP: why not showing normally distributed variations (before and after calibration)? TP: good idea, will do, if some time left}

\citet{Brunel00} reports \emph{balanced random networks} (BRNs) exhibiting, among others, asynchronous irregular network states with stationary global activity.

\subsubsection{\network{}}

BRNs consist of an inhibitory and excitatory population of neurons, both receiving feedforward connections from two populations of Poisson processes mimicking background activity.
Both neuron populations are recurrently connected including connections within the populations.
All connections are realized with random and sparse connections of probability $p$.
In this study, synaptic weights for inhibitory connections are chosen four times larger than those for excitatory ones.
In contrast to the original implementation using $12500$ neurons, we scaled this network by a factor of $100$ while preserving its firing behavior.

If single cells fire irregularly, the \emph{coefficient of variation}
\begin{equation}
\cv=\frac{\mrmsub{\sigma}{T}}{\overline{T}}
\end{equation}
of interspike intervals has values close to or higher than one \citep{Dayan01}.
$\overline{T}$ and $\mrmsub{\sigma}{T}$ are the mean and standard deviation of these intervals.
Synchrony between two cells can be measured by calculating the \emph{correlation coefficient}
\begin{equation}
\cc=\frac{\mathrm{cov}(n_1, n_2)}{\sqrt{\mathrm{var}(n_1) \mathrm{var}(n_2)}}
\end{equation}
of their spike trains $n_1$ and $n_2$, respectively \citep{Perkel67b}.
The variance ($\mathrm{var}$) and covariance ($\mathrm{cov}$) are calculated by using time bins with \SI{2}{\milli\second} duration \citep{Kumar08_1}.

\citet{Bruederle10_iscas} have shown another approach to investigate networks inspired by \citet{Brunel00}.
Their focus have been the effects of network parameters and STP on the firing rate of the network.
In our study, we show that such BRNs can show an asynchronous irregular network state, when emulated on hardware.

\subsubsection{\hardware{}}

\ifdraftmode
\begin{figure*}[tb]
\ifdraftmode
  \mbox{
  \hspace{-2.25cm}
\fi
\includegraphics[width=\fullfigwidth]{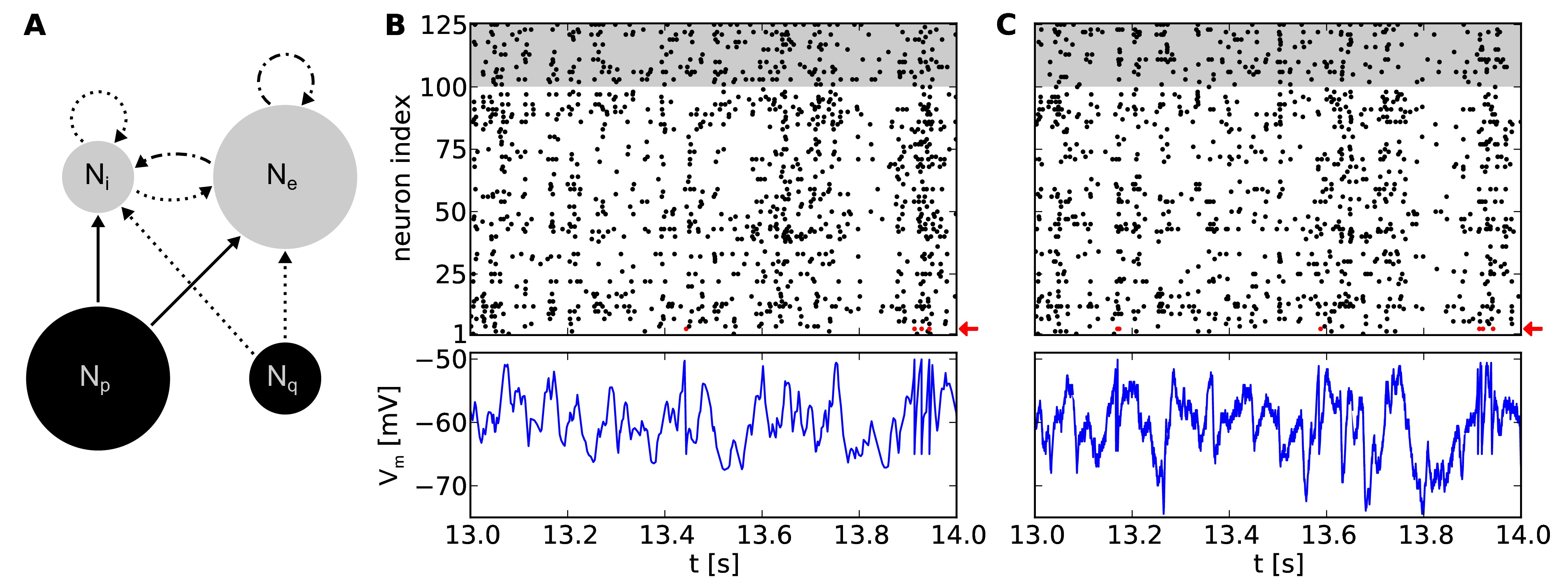}
\ifdraftmode
  }
\fi
\caption{
\textbf{(A)}
Network topology of a balanced random network.
Populations consisting of $\mrmsub{N}{e}=100$ excitatory and $\mrmsub{N}{i}=25$ inhibitory neurons (gray circles), respectively, are stimulated by populations of Poisson sources (black circles).
We use $\mrmsub{N}{p}=100$ independent sources for excitation and $\mrmsub{N}{q}=25$ for inhibition.
Arrows denote projections between these populations with connection probabilities $p=0.1$, with solid lines for excitatory and dotted lines for inhibitory connections.
Dot and dash lines are indicating excitatory projections with short-term depression.
\textbf{(B)}
Top:
Raster plot of a software simulation.
Populations of excitatory and inhibitory neurons are depicted with white and gray background, respectively.
Note that for clarity only the time interval [\SI{13}{\second},~\SI{14}{\second}] of a \SI{20}{\second} emulation is shown.
For the full \SI{20}{\second} emulation, we have measured $\cv=0.96\pm0.09$ (mean over all neurons) and $\cc=0.010\pm0.017$ (mean over 1000 random chosen pairs of neurons), respectively.
Bottom:
Recorded membrane potential of an arbitrary excitatory neuron (neuron index 3, highlighted with a red arrow in the above raster plot).
\textbf{(C)}
Same network topology and stimulus as in (B), but emulated on the \spikey{} chip, resulting in $\cv=1.02\pm0.16$ and $\cc=0.014\pm0.019$.
Note that the membrane recordings are calibrated such that the threshold and reset potential match those of the software counterpart.
\label{fig:brn}}
\end{figure*}
\fi

In addition to standard calibration routines (\prettyref{sec:hw_calibration}), we have calibrated the chip explicitly for the BRN shown in \prettyref{fig:brn}A.
In the first of two steps, excitatory and inhibitory synapse line drivers were calibrated sequentially towards equal strength, respectively, but with inhibition four times stronger than excitation.
To this end, all available neurons received spiking activity from a single synapse line driver, thereby averaging out neuron-to-neuron variations.
The shape of synaptic conductances (specifically $\tfall$ and $\gimax$) were adjusted to obtain a target mean firing rate of \SI{10}{\hertz} over all neurons.
Similarly, each driver was calibrated for its inhibitory operation mode.
All neurons were strongly stimulated by an additional driver with its excitatory mode already calibrated, and again the shape of conductances, this time for inhibition, was adjusted to obtain the target rate.

Untouched by this prior calibration towards a target mean rate, neuron excitability still varied between neurons and was calibrated consecutively for each neuron in a second calibration step.
For this, all neurons of the BRN were used to stimulate a single neuron with a total firing rate that was uniformly distributed among all inputs and equal to the estimated firing rate of the final network implementation.
Subsequently, all afferent synaptic weights to this neuron were scaled in order to adapt its firing rate to the target rate.

To avoid a self-reinforcement of network activity observed in emulations on the hardware, efferent connections of the excitatory neuron population were modeled as short-term depressing.
Nevertheless, such BRNs still show an asynchronous irregular network state (\prettyref{fig:brn}B).

\prettyref{fig:brn}C show recordings of a BRN emulation on a calibrated chip with neurons firing irregularly and asynchronously.
Note that $\cv \ge 1$ does not necessarily guarantee an exponential interspike interval distribution and even less Poisson firing.
However, neurons within the BRN clearly exhibit irregular firing (compare raster plots of \prettyref{fig:brn}B and C).

A simulation of the same network topology and stimulus using software tools produced similar results.
Synaptic weights were not known for the hardware emulation, but defined by the target firing rates using the above calibration.
A translation to biological parameters is possible, but would have required further measurements and was not of further interest in this context.
Instead, for software simulations, the synaptic weight for excitatory connections were chosen to fit the mean firing rate of the hardware emulation (approx. \SI{9}{\hertz}).
Then, the weight of inhibitory connections were chosen to preserve the ratio between inhibitory and excitatory weights.

Membrane dynamics of single neurons within the network are comparable between hardware emulations and software simulations (\prettyref{fig:brn}B and C).
Evidently, spike times differ between the two approaches due to various hardware noise sources (\prettyref{sec:hw_calibration}).
However, in ``large'' populations of neurons (${N}{e} + {N}{i} = 125$ neurons), we observe that these phenomena have qualitatively no effect on firing statistics, which are comparable to software simulations (compare raster plots of \prettyref{fig:brn}B and C).
The ability to reproduce these statistics is highly relevant in the context of cortical models which rely on asynchronous irregular firing activity for information processing \citep[e.g.,][]{Vreeswijk96}.
\subsection{Soft Winner-Take-All Network}
\label{sec:wta}

\ifdraftmode
\else
\begin{figure*}[tbp]
\ifdraftmode
  \mbox{
  \hspace{-2.25cm}
\fi
\includegraphics[width=\fullfigwidth]{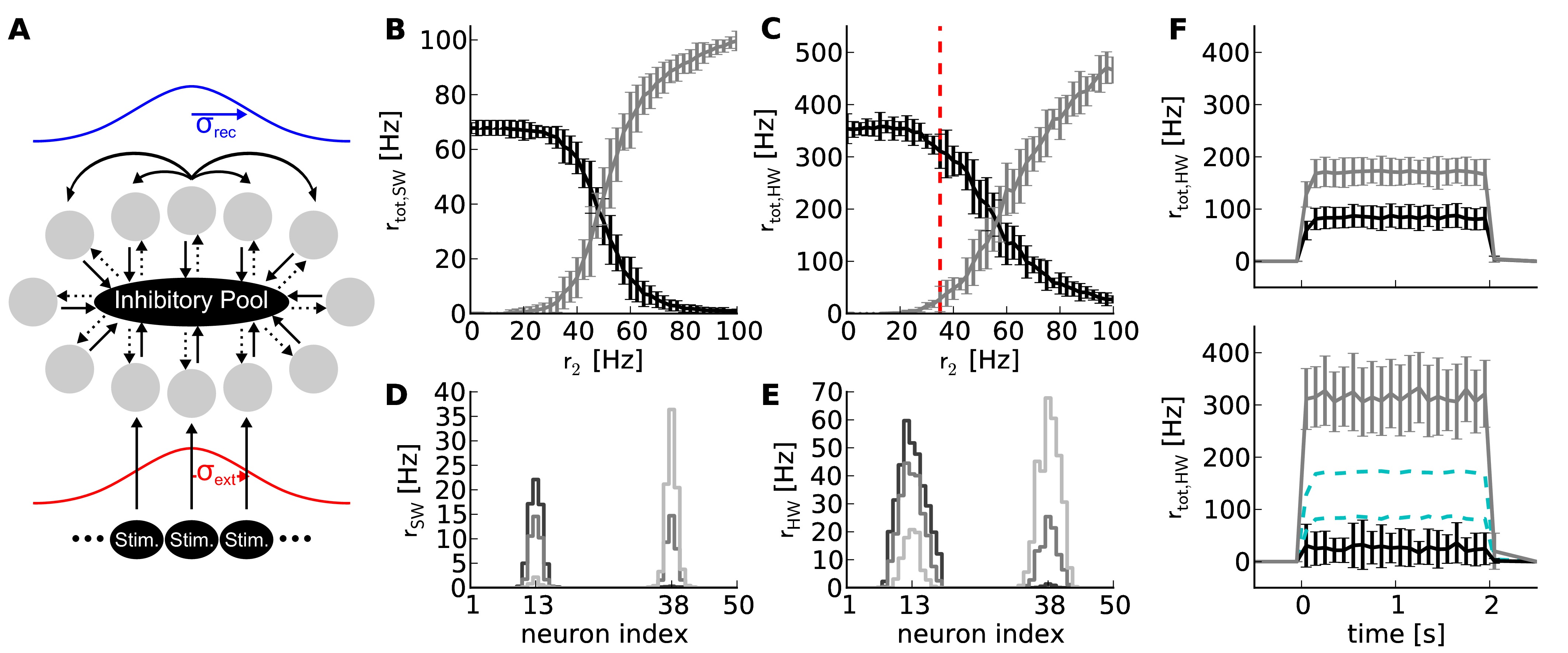}
\ifdraftmode
  }
\fi
\caption{
\textbf{(A)}
Topology of a soft winner-take-all network with 50 excitatory (gray circles) and 16 inhibitory neurons.
Solid and dotted arrows denote excitatory and inhibitory connections, respectively.
The strength profile of recurrent connections between excitatory neurons and external stimulations is schematized in blue and red, respectively (for details see text).
All projections between neuron populations have a connection probabilities of $p=1$, except the projection between the excitatory and inhibitory neuron population ($p=0.6$).
\textbf{(B)}
Results of software simulation (SW). 
Black curve: Total firing rate of the reference half where constant external stimulation is received ($r_1=\SI{50}{\hertz}$ at $\mrmsub{\mu}{ext}=\text{neuron index }13$).
Gray curve: Total firing rate of the neurons in the half of the ring where varying external stimulation with rate $r_2$ between zero and \SI{100}{\hertz} is received (at $\mrmsub{\mu}{ext}=\text{neuron index }38$). 
Firing rates $\mrmsub{r}{tot}$ of all neurons in each half of the ring were averaged over $10$ runs with \SI{2}{\second} duration and different random number seeds for drawing the stimulus spike trains.
\textbf{(C)}
Same network topology and stimulus as (B), but emulated on \spikey{} (HW).
\textbf{(D)}
Firing rate distribution over neuron indices for $r_2=\SI{25}{\hertz}$ (black), \SI{50}{\hertz} (dark gray) and \SI{75}{\hertz} (light gray).
\textbf{(E)}
Same as (D), but emulated on \spikey{}.
\textbf{(F)}
Top panel: Time course of firing rates for stimulus indicated in (C, red dashed line), but without recurrent connections.
All excitatory neurons are solely driven by an external stimulus of $r_1=\SI{50}{\hertz}$ and $r_2=\SI{35}{\hertz}$, respectively.
Firing rates were averaged over $100$ runs.
Bottom panel: Same as top panel, but with recurrent connections. For better comparison, data of the top panel is drawn in cyan dashed lines.
\label{fig:wta}}
\end{figure*}
\fi

Soft winner-take-all (sWTA) computation is often viewed as an underlying principle in models of cortical processing \citep{grossberg73_213,Maass00_2519,Itti2001_194,Douglas04,Oster09_2437,Lundqvist10_e803}. 
The sWTA architecture has many practical applications, for example contrast enhancement, or making a decision which of two concurrent inputs is larger.
Many neuromorphic systems explicitly implement sWTA architectures \citep{Lazzaro88_703, Chicca07_981, Neftci11_2457}.

\subsubsection{\network{}}

We implemented an sWTA network that is composed of a ring-shaped layer of recurrently connected excitatory and a common pool of inhibitory neurons (\prettyref{fig:wta}A), following the implementation by \citet{Neftci11_2457}.
Excitatory neurons project to the common inhibitory pool and receive recurrent feedback from there. 
In addition, excitatory neurons have recurrent excitatory connections to their neighbors on the ring. 
The strength of these decays with increasing distance on the ring, following a Gaussian profile with a standard deviation of $\mrmsub{\sigma}{rec} = 5$ neurons.
External stimulation is also received through a Gaussian profile, with the mean $\mrmsub{\mu}{ext}$ expressing the neuron index that receives input with maximum synaptic strength. 
Synaptic input weights to neighbors of that neuron decay according to a standard deviation of $\mrmsub{\sigma}{ext} = 3$ neurons.
We clipped the input weights to zero beyond $\mrmsub{\sigma}{ext} \cdot 3$.
Each neuron located within the latter Gaussian profile receives stimulation from five independent Poisson spike sources each firing at rate $r$.
Depending on the contrast between the input firing rates $r_1$ and $r_2$ of two stimuli applied to opposing sides of the ring, one side of the ring ``wins'' by firing with a higher rate and thereby suppressing the other.

\ifdraftmode
\begin{figure*}[tbp]
\ifdraftmode
  \mbox{
  \hspace{-2.25cm}
\fi
\includegraphics[width=\fullfigwidth]{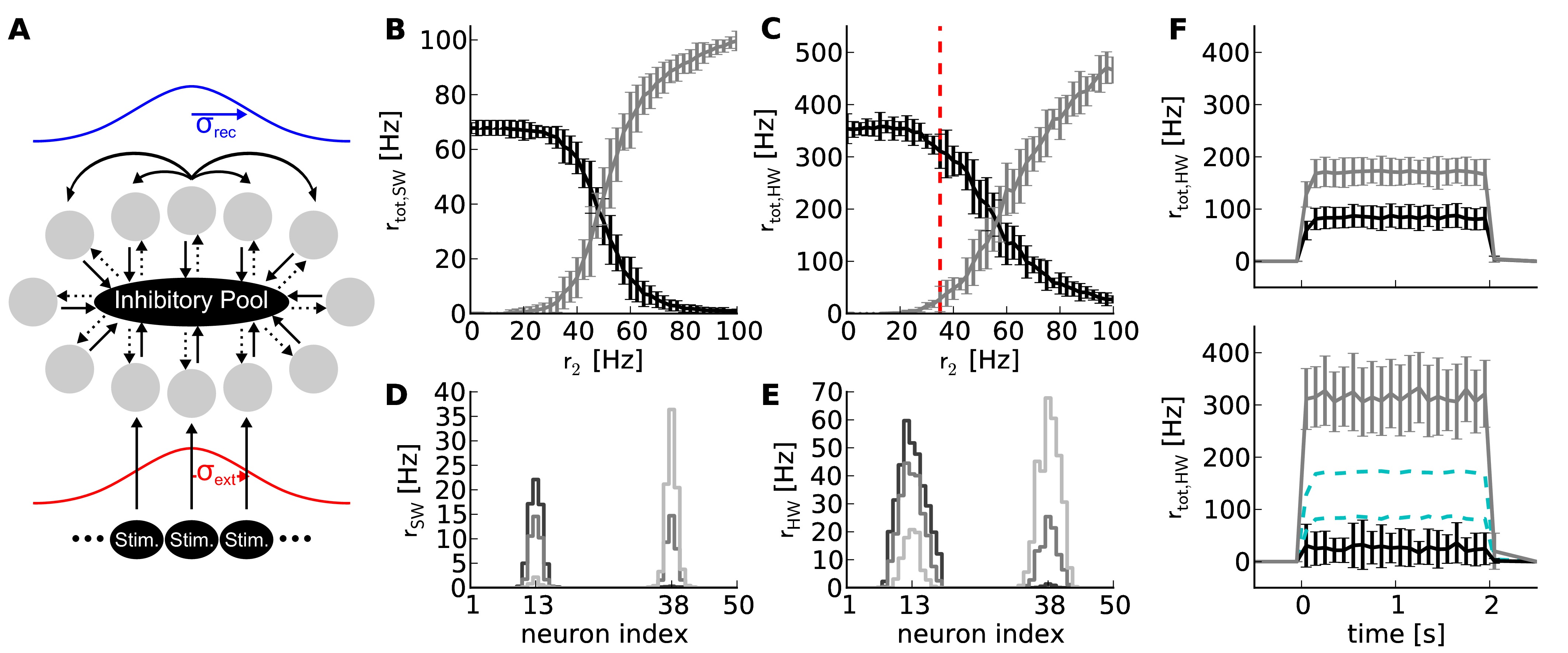}
\ifdraftmode
  }
\fi
\caption{
\textbf{(A)}
Topology of a soft winner-take-all network with 50 excitatory (gray circles) and 16 inhibitory neurons.
Solid and dotted arrows denote excitatory and inhibitory connections, respectively.
The strength profile of recurrent connections between excitatory neurons and external stimulations is schematized in blue and red, respectively (for details see text).
All projections between neuron populations have a connection probabilities of $p=1$, except the projection between the excitatory and inhibitory neuron population ($p=0.6$).
\textbf{(B)}
Results of software simulation (SW). 
Black curve: Total firing rate of the reference half where constant external stimulation is received ($r_1=\SI{50}{\hertz}$ at $\mrmsub{\mu}{ext}=\text{neuron index }13$).
Gray curve: Total firing rate of the neurons in the half of the ring where varying external stimulation with rate $r_2$ between zero and \SI{100}{\hertz} is received (at $\mrmsub{\mu}{ext}=\text{neuron index }38$). 
Firing rates $\mrmsub{r}{tot}$ of all neurons in each half of the ring were averaged over $10$ runs with \SI{2}{\second} duration and different random number seeds for drawing the stimulus spike trains.
\textbf{(C)}
Same network topology and stimulus as (B), but emulated on \spikey{} (HW).
\textbf{(D)}
Firing rate distribution over neuron indices for $r_2=\SI{25}{\hertz}$ (black), \SI{50}{\hertz} (dark gray) and \SI{75}{\hertz} (light gray).
\textbf{(E)}
Same as (D), but emulated on \spikey{}.
\textbf{(F)}
Top panel: Time course of firing rates for stimulus indicated in (C, red dashed line), but without recurrent connections.
All excitatory neurons are solely driven by an external stimulus of $r_1=\SI{50}{\hertz}$ and $r_2=\SI{35}{\hertz}$, respectively.
Firing rates were averaged over $100$ runs.
Bottom panel: Same as top panel, but with recurrent connections. For better comparison, data of the top panel is drawn in cyan dashed lines.
\label{fig:wta}}
\end{figure*}
\fi

\subsubsection{\hardware{}}
We assessed the efficiency of this sWTA circuit by measuring the reduction in firing rate exerted in neurons when the opposite side of the ring is stimulated.
We stimulated one side of the ring with a constant, and the opposite side with a varying firing rate.
In case of hardware emulations, each stimulus was distributed and hence averaged over multiple line drivers in order to equalize stimulation strength among neurons.
For both back-ends, inhibitory weights were chosen four times stronger than excitatory ones (using the synapse line driver calibration of \prettyref{sec:brn}).

The firing rate of the reference side decreased when the firing rate of stimulation to the opposite side was increased, both in software simulation and on the hardware (\prettyref{fig:wta}B and C). 
In both cases, the average firing rates crossed at approximately $r_2 = \SI{50}{\hertz}$, corresponding to the spike rate delivered to the reference side. 
The firing rates $\mrmsub{r}{tot}$ are less distinctive for hardware emulations compared to software simulations, but still sufficient to produce robust sWTA functionality.
Note that the observed firing rates are higher on the hardware than in the software simulation. 
This difference is due to the fact that the reliability of the network performance improved for higher firing rates.

\prettyref{fig:wta}D and E depict activity profiles of the excitatory neuron layer.
The hardware neurons exhibited a broader and also slightly asymmetric excitation profile compared to the software simulation. 
The asymmetry is likely due to inhomogeneous excitability of neurons, which is caused by fixed-pattern noise (\prettyref{sec:hw_system}). 
The broader excitation profile indicates that inhibition is less efficient on the hardware than in the software simulation (a trend that can also be observed in the firing rates in \prettyref{fig:wta}B and C). 
Counteracting this loss of inhibition may be possible through additional calibration, if the sharpness of the excitation profile is critical for the task in which such an sWTA circuit is to be employed.

The network emulated on \spikey{} is said to perform sWTA, because the side of the ring with stronger stimulation shows an amplified firing rate, while the firing rate of the other side is suppressed (see \prettyref{fig:wta}F).
This qualifies our hardware system for applications relying on similar sWTA network topologies.
\subsection{Cortical Layer 2/3 Attractor Model}
\label{sec:kth}

Throughout the past decades, attractor networks that model working memory in the cerebral cortex have gained increasing support from both experimental data and computer simulations.
The \emph{cortical layer 2/3 attractor memory model} described in \citet{Lundqvist06_253,Lundqvist10_e803} has been remarkably successful at reproducing both low-level (firing patterns, membrane potential dynamics) and high level (pattern completion, attentional blink) features of cortical information processing.
One particularly valuable aspect is the very low amount of fine-tuning this model requires in order to reproduce the rich set of desired internal dynamics.
It has also been shown in \citet{Bruederle11_263} that there are multiple ways of scaling this model down in size without affecting its main functionality features.
These aspects make it an ideal candidate for implementation on our analog neuromorphic device.
In this context, it becomes particularly interesting to analyze how the strong feedback loops which predominantly determine the characteristic network activity are affected by the imposed limitations of the neuromorphic substrate and fixed-pattern noise.
Here, we extend the work done in \citet{Bruederle11_263} by investigating specific attractor properties such as firing rates, voltage UP-states and the pattern completion capability of the network.

\subsubsection{\network{}}

\ifdraftmode
\else
\ifdraftmode
  \begin{figure*}[tbp]
\else
  \begin{figure*}[tb]
\fi

\ifdraftmode
  \mbox{
  \hspace{-2.25cm}
\fi
\includegraphics[width=\fullfigwidth]{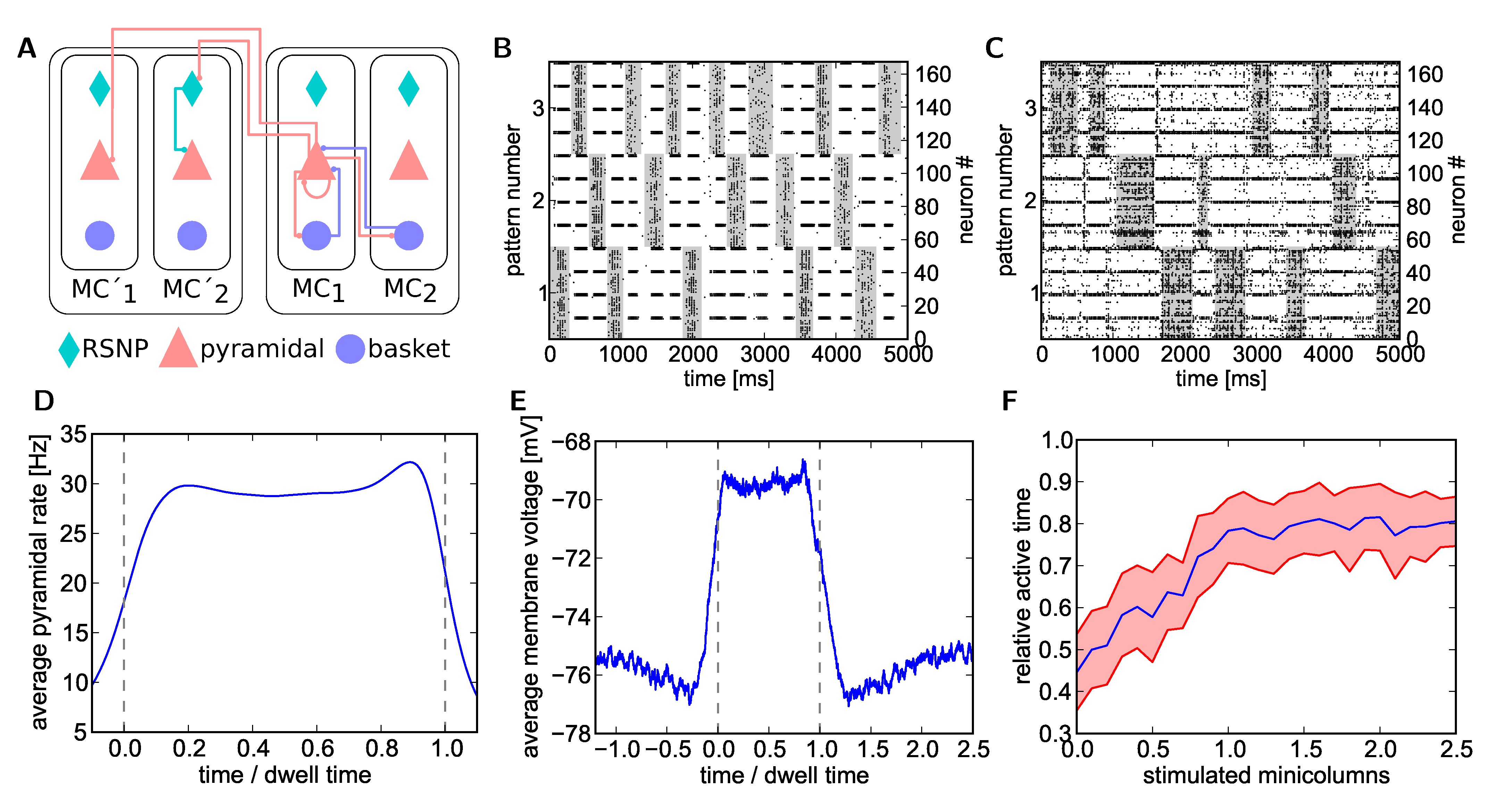}
\ifdraftmode
  }
\fi
\caption{
\textbf{(A)}
Schematic of the cortical layer 2/3 attractor memory network.
Two hypercolumns, each containing two minicolumns, are shown.
For better readability, only connections that are active within an active pattern are depicted.
See text for details.
\textbf{(B)}
Software simulation of spiking activity in the cortical attractor network model scaled down to 192 neurons (only pyramidal and RSNP cells shown, basket cells spike almost continously).
Minicolumns belonging to the same pattern are grouped together.
The broad stripes of activity are generated by pyramidal cells in active attractors.
The interlaced narrow stripes of activity represent pairs of RSNP cells, which spike when their home minicolumn is inhibited by other active patterns.
\textbf{(C)}
Same as \textbf{B}, but on hardware.
The raster plot is noisier and the duration of attractors (dwell time) are less stable than in software due to fixed-pattern noise on neuron and synapse circuits.
For better readability, active states are underlied in grey in \textbf{B} and \textbf{C}.
\textbf{(D)}
Average firing rate of pyramidal cells on the \spikey{} chip inside active patterns.
To allow averaging over multiple active periods of varying lengths, all attractor dwell times have been normalized to $1$.
\textbf{(E)}
Average membrane potential of pyramidal cells on the \spikey{} chip inside and outside active patterns.
\textbf{(F)}
Pattern completion on the \spikey{} chip.
Average values (from multiple runs) depicted in blue, with the standard deviation shown in red.
From a relatively equilibrated state where all patterns take turns in being active, additional stimulation (see text) of only a subset of neurons from a given attractor activates the full pattern and enables it to dominate over the other two.
The pattern does not remain active indefinitely due to short-term depression in excitatory synapses, thereby still allowing short occasional activations of the other two patterns.
\label{fig:l23}}
\smalltodominor[inline]{TP: Blobs in A too small?!}
\end{figure*}
\fi

From a structural perspective, the most prominent feature of the Layer 2/3 Attractor Memory Network is its modularity.
Faithful to its biological archetype, it implements a set of cortical hypercolumns, which are in turn subdivided into multiple minicolumns (\prettyref{fig:l23}A).
Each minicolumn consists of three cell populations: excitatory pyramidal cells, inhibitory basket cells and inhibitory RSNP (regular spiking non-pyramidal) cells.

Attractor dynamics arise from the synaptic connectivity on two levels.
Within a hypercolumn, the basket cell population enables a soft-WTA-like competition among the pyramidal populations within the minicolumns.
On a global scale, the long-range inhibition mediated by the RSNP cells governs the competition among so-called \emph{patterns}, as explained in the following.

In the original model described in \citet{Lundqvist10_e803}, each hypercolumn contains 9 minicolumns, each of which consists of 30 pyramidal, 2 RSNP and 1 basket cells.
Within a minicolumn, the pyramidal cells are interconnected and also project onto the 8 closest basket cells within the same hypercolumn.
In turn, pyramidal cells in a minicolumn receive projections from all basket cells within the same hypercolumn.
All pyramidal cells receive two types of additional excitatory input: an evenly distributed amount of diffuse Poisson noise and specific activation from the cortical layer 4.
Therefore, the minicolumns (i.e., the pyramidal populations within) compete among each other in WTA-like fashion, with the winner being determined by the overall strength of the received input.

A pattern (or attractor) is defined as containing exactly one minicolumn from each hypercolumn.
Considering only orthogonal patterns (each minicolumn may only belong to a single pattern) and given that all hypercolumns contain an equal amount of minicolumns, the number of patterns in the network is equal to the number of minicolumns per hypercolumn.
Pyramidal cells within each minicolumn project onto the pyramidal cells of all the other minicolumns in the same pattern.
These connections ensure a spread of local activity throughout the entire pattern.
Additionally, the pyramidal cells also project onto the RSNP cells of all minicolumns belonging to different attractors, which in turn inhibit the pyramidal cells within their minicolumn.
This long-range competition enables the winning pattern to completely shut down the activity of all other patterns.

Two additional mechanisms weaken active patterns, thereby facilitating switches between patterns.
The pyramidal cells contain an adaptation mechanism which decreases their excitability with every emitted spike.
Additionally, the synapses between pyramidal cells are modeled as short-term depressing.

\ifdraftmode
\ifdraftmode
  \begin{figure*}[tbp]
\else
  \begin{figure*}[tb]
\fi

\ifdraftmode
  \mbox{
  \hspace{-2.25cm}
\fi
\includegraphics[width=\fullfigwidth]{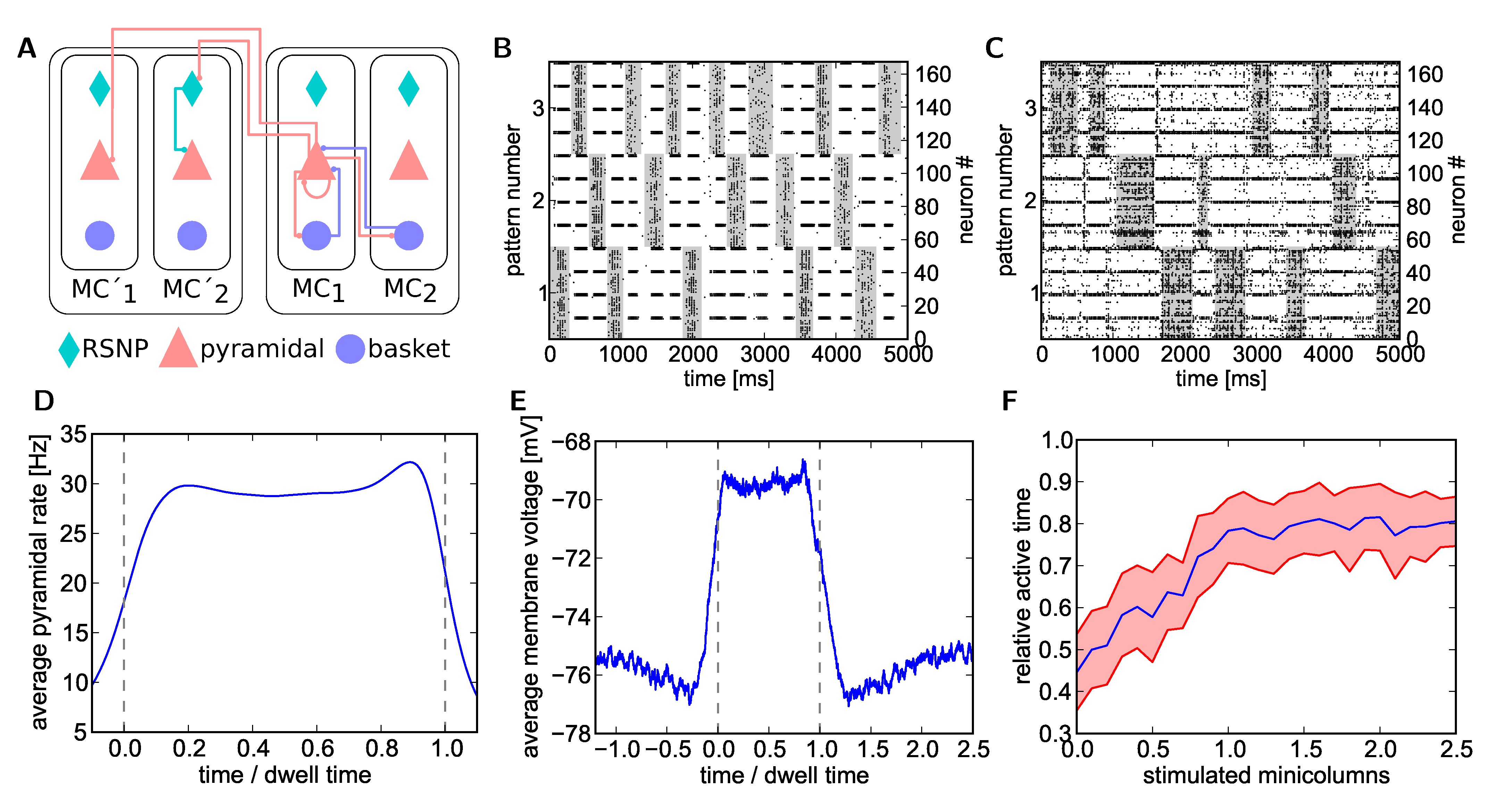}
\ifdraftmode
  }
\fi
\caption{
\textbf{(A)}
Schematic of the cortical layer 2/3 attractor memory network.
Two hypercolumns, each containing two minicolumns, are shown.
For better readability, only connections that are active within an active pattern are depicted.
See text for details.
\textbf{(B)}
Software simulation of spiking activity in the cortical attractor network model scaled down to 192 neurons (only pyramidal and RSNP cells shown, basket cells spike almost continously).
Minicolumns belonging to the same pattern are grouped together.
The broad stripes of activity are generated by pyramidal cells in active attractors.
The interlaced narrow stripes of activity represent pairs of RSNP cells, which spike when their home minicolumn is inhibited by other active patterns.
\textbf{(C)}
Same as \textbf{B}, but on hardware.
The raster plot is noisier and the duration of attractors (dwell time) are less stable than in software due to fixed-pattern noise on neuron and synapse circuits.
For better readability, active states are underlied in grey in \textbf{B} and \textbf{C}.
\textbf{(D)}
Average firing rate of pyramidal cells on the \spikey{} chip inside active patterns.
To allow averaging over multiple active periods of varying lengths, all attractor dwell times have been normalized to $1$.
\textbf{(E)}
Average membrane potential of pyramidal cells on the \spikey{} chip inside and outside active patterns.
\textbf{(F)}
Pattern completion on the \spikey{} chip.
Average values (from multiple runs) depicted in blue, with the standard deviation shown in red.
From a relatively equilibrated state where all patterns take turns in being active, additional stimulation (see text) of only a subset of neurons from a given attractor activates the full pattern and enables it to dominate over the other two.
The pattern does not remain active indefinitely due to short-term depression in excitatory synapses, thereby still allowing short occasional activations of the other two patterns.
\label{fig:l23}}
\smalltodominor[inline]{TP: Blobs in A too small?!}
\end{figure*}
\fi

\subsubsection{\hardware{}}

When scaling down the original model (2673 neurons) to the maximum size available on the \spikey{} chip (192 neurons,  see \prettyref{fig:l23}B for software simulation results), we made use of the essential observation that the number of pyramidal cells can simply be reduced without compensating for it by increasing the corresponding projection probabilities.
Also, for less than 8 minicolumns per hypercolumn, all basket cells within a hypercolumn have identical afferent and efferent connectivity patterns, therefore allowing to treat them as a single population.
Their total number was decreased, while increasing their efferent projection probabilities accordingly.
In general (i.e., except for pyramidal cells), when number and/or size of populations were changed, projection probabilities were scaled in such a way that the total fan-in for each neuron was kept at a constant average.
When the maximum fan-in was reached (one afferent synapse for every neuron in the receptive field), the corresponding synaptic weights were scaled up by the remaining factor.

Because neuron and synapse models on the \spikey{} chip are different to the ones used in the original model, we have performed a heuristic fit in order to approximately reproduce the target firing patterns.
Neuron and synapse parameters were first fitted in such a way as to generate clearly discernible attractors with relatively high average firing rates (see \prettyref{fig:l23}D).
Additional tuning was needed to compensate for missing neuronal adaptation, limitations in hardware configurability, parameter ranges and fixed-pattern noise affecting hardware parameters.

During hardware emulations, apart from the appearance of spontaneous attractors given only diffuse Poisson stimulation of the network (\prettyref{fig:l23}C), we were able to observe two further interesting phenomena which are characteristic for the original attractor model.

When an attractor becomes active, its pyramidal cells enter a so-called UP state which is characterized by an elevated average membrane potential.
\prettyref{fig:l23}E clearly shows the emergence of such UP-states on hardware.
The onset of an attractor is characterized by a steep rise in pyramidal cell average membrane voltage, which then decays towards the end of the attractor due to synaptic short-term depression and/or competition from other attractors temporarily receiving stronger stimulation.
On both flanks of an UP state, the average membrane voltage shows a slight undershoot, due to the inhibition by other active attractors.

A second important characteristic of cortical attractor models is their capability of performing \emph{pattern completion} \citep{Lundqvist06_253}.
This means that a full pattern can be activated by stimulating only a subset of its constituent pyramidal cells (in the original model, by cells from cortical Layer 4, modeled by us as additional Poisson sources).
The appearance of this phenomenon is similar to a phase transition from a resting state to a collective pyramidal UP-state occurring when a critical amount of pyramidal cells are stimulated.
To demonstrate pattern completion, we have used the same setup as in the previous experiments, except for one pattern receiving additional stimulation.
From an initial equilibrium between the three attractors (approximately equal active time), we have observed the expected sharp transition to a state where the stimulated attractor dominates the other two, occurring when one of its four minicolumns received L4 stimulus (\prettyref{fig:l23}F).

The implementation of the attractor memory model is a particularly comprehensive showcase of the configurability and functionality of our neuromorphic platform due to the complexity of both model specifications and emergent dynamics.
Starting from these results, the next-generation hardware \citep{Schemmel10_1947} will be able to much more accurately model biological behavior, thanks to a more flexible, adapting neuron model and a significantly increased network size.
\subsection{Insect Antennal Lobe Model}
\label{sec:al}

\ifdraftmode
\else
\begin{figure*}[tb]
 \ifdraftmode
  \mbox{
  \hspace{-2.25cm}
 \fi
 \includegraphics[width=\fullfigwidth]{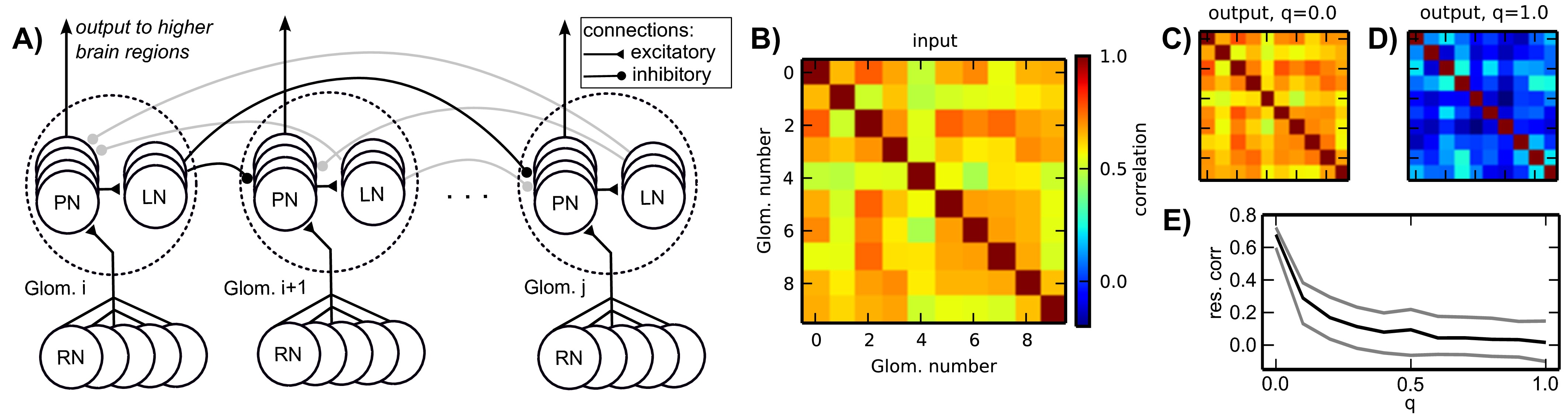}
 \ifdraftmode
  }
 \fi
 \caption{
\textbf{(A)}
Schematic of the insect antennal lobe network. 
Neuron populations are grouped in glomeruli (outlined by dotted lines), which exert lateral inhibition onto each other. 
RNs: receptor neurons (input), PNs: projection neurons (output), LNs: inhibitory local neurons. 
Some connections are grayed out to emphasize the connection principle.
\textbf{(B)}
Correlation matrix of the input data.
\textbf{(C)}
Correlation matrix of the output spike rates (PNs) without lateral inhibition, $q=0.0$.
\textbf{(D)}
Correlation of the output with homogeneous lateral inhibition, $q=1.0$.
\textbf{(E)}
Average pairwise
correlation between glomeruli (median $\pm$ $20^{th}$ (black) and $80^{th}$ (gray) percentile)
in dependence of the overall strength of lateral inhibition $q$. 
\label{fig:insectAL}}
\smalltodominor[inline]{TP: very minor: black-white color plots}
\end{figure*}
\fi

The high acceleration factor of the \spikey{} chip makes it an attractive platform for neuromorphic data processing.
Preprocessing of multivariate data is a common problem in signal and data analysis. 
In conventional computing, reduction of correlation between input channels is often the first step in the analysis of multidimensional data, achieved, e.g., by \emph{principal component analysis} (PCA).
The architecture of the olfactory system maps particularly well onto this problem \citep{Schmuker_Schneider_2007}. 
We have implemented a network that is inspired by processing principles that have been described in the insect
antennal lobe (AL), the first relay station from olfactory sensory neurons to higher brain areas. 
The function of the AL has been described to decorrelate the inputs from sensory neurons, potentially enabling more efficient memory formation and retrieval \citep{Stopfer_Bhagavan_Smith_Laurent_1997, Linster_Smith_1997,
Perez-Orive_Bazhenov_Laurent_2004, Wilson_Laurent_2005}. The mammalian analog of the AL (the olfactory bulb) has been the target of a recent neuromorphic modeling study \citep{Imam12_83}.

The availability of a network building block that achieves channel decorrelation is an important step toward high-performance neurocomputing. 
The aim of this experiment is to demonstrate that the previously studied rate-based AL model \citep{Schmuker_Schneider_2007} that reduces rate correlation between input channels is applicable to a spiking neuromorphic hardware system.

\subsubsection{\network{}}

In the insect olfactory system, odors are first encoded into neuronal signals by receptor neurons (RNs) which are located on the antenna.
RNs send their axons to the AL (\prettyref{fig:insectAL}A).
The AL is composed of glomeruli, spherical compartments where RNs project onto local inhibitory neurons (LNs) and projection neurons (PNs). 
LNs project onto other glomeruli, effecting lateral inhibition. 
PNs relay the information to higher brain areas where multimodal integration and memory formation takes place. 

The architecture of our model reflects the neuronal connectivity in the insect
AL (\prettyref{fig:insectAL}A). RNs are modeled as spike train
generators,
which project onto the 
PNs in the corresponding glomerulus. The PNs project onto the LNs, which send inhibitory projections to the PNs in other glomeruli.

In biology, the AL network reduces the rate correlation between glomeruli, in order to improve stimulus separability and thus odor identification. 
Another effect of decorrelation is that the rate patterns encoding the stimuli become sparser, and use the available coding space more efficiently as redundancy is reduced. 
Our goal was to demonstrate the reduction of rate correlations across glomeruli (\emph{channel correlation}) by the AL-inspired spiking network.
To this end, we generated patterns of firing rates with channel correlation. 
We created a surrogate data set exhibiting channel correlation using a copula, a technique that allows to generate correlated series of samples from an arbitrary random distribution and a covariance matrix \citep{Nelsen_1998}. 
The covariance matrix was uniformly set to a target correlation of $0.6$.
Using this copula, we sampled 100 ten-dimensional data vectors from an exponential distribution. 
In the biological context, this is equivalent to having a repertoire of 100 odors, each encoded by ten receptors, and the firing rate of each input channel following a decaying exponential distribution.
Values larger than $e$ were clipped and the distribution was mapped to the interval $[0,1]$ by applying $v = v/e$ for each value $v$. 
These values were then converted into firing rates between 20 and 55 spikes/s. 
The ten-dimensional data vector was presented to the network by mapping the ten firing rates onto the ten glomeruli, setting all single RNs in each glomerulus to fire at the respective target rates.
Rates were converted to spike trains individually for each RN using the Gamma process with $\gamma=5$. 
Each data vector was presented to the network for the duration of one second by making the RNs of each glomerulus fire with the specified rate. 
The inhibitory weights between glomeruli were uniform, i.e., all inhibitory connections shared the same weight.
During one second of stimulus presentation, output rates were measured from PNs. 
One output rate per glomerulus was obtained by averaging the firing rate of all PNs in a glomerulus. 

We have used 6 RN input streams per glomerulus, projecting in an all-to-all fashion onto 7 PNs, which in turn projected on 3 LNs per glomerulus.

\ifdraftmode
\begin{figure*}[tb]
 \ifdraftmode
  \mbox{
  \hspace{-2.25cm}
 \fi
 \includegraphics[width=\fullfigwidth]{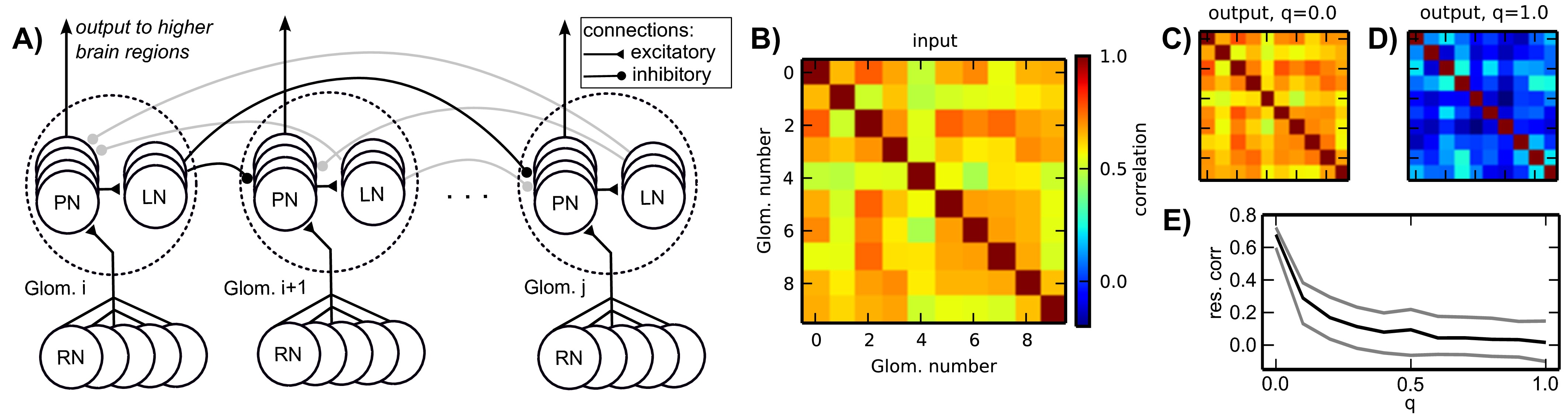}
 \ifdraftmode
  }
 \fi
 \caption{
\textbf{(A)}
Schematic of the insect antennal lobe network. 
Neuron populations are grouped in glomeruli (outlined by dotted lines), which exert lateral inhibition onto each other. 
RNs: receptor neurons (input), PNs: projection neurons (output), LNs: inhibitory local neurons. 
Some connections are grayed out to emphasize the connection principle.
\textbf{(B)}
Correlation matrix of the input data.
\textbf{(C)}
Correlation matrix of the output spike rates (PNs) without lateral inhibition, $q=0.0$.
\textbf{(D)}
Correlation of the output with homogeneous lateral inhibition, $q=1.0$.
\textbf{(E)}
Average pairwise
correlation between glomeruli (median $\pm$ $20^{th}$ (black) and $80^{th}$ (gray) percentile)
in dependence of the overall strength of lateral inhibition $q$. 
\label{fig:insectAL}}
\smalltodominor[inline]{TP: very minor: black-white color plots}
\end{figure*}
\fi

\subsubsection{\hardware{}}

The purpose of the presented network was to reduce rate correlation between input channels. 
As in other models, fixed-pattern noise across neurons had a detrimental effect on the function of the network. 
We exploited the specific structure of our network to implement more efficient calibration than can be provided by standard calibration methods (\prettyref{sec:hw_calibration}).
Our calibration algorithm targeted PNs and LNs in the first layer of the network. 
During calibration, we turned off all projections between glomeruli.
Its aim was to achieve a homogeneous response across PNs and LNs respectively, i.e., within $\pm$ 10\% of a target rate. 
The target rate was chosen from the median response rate of uncalibrated neurons. 
For neurons whose response rate was too high it was sufficient to reduce the synaptic weight of the excitatory input from RNs. 
For those neurons with a too low rate the input strength had to be increased.
The excitatory synaptic weight of the input from RNs was initially already at its maximum value and could not be increased. 
As a workaround we used PNs from the same glomerulus to add additional excitatory input to those ``weak'' neurons.
We ensured that no recurrent excitatory loops were introduced by this procedure.
If all neurons in a glomerulus were too weak, we recruit another external input stream to achieve the desired target rate. 
Once the PNs were successfully calibrated (less than 10\% deviation from the target rate), we used the same approach to calibrate the LNs in each glomerulus.

To assess the performance of the network we have compared the channel correlation in the input and in the output.
The channel correlation matrix $\mathbf{C}$ was computed according to
\begin{equation}
\label{eq:channel_correlation} 
C_{i,j} = d^{\text{Pearson}}({\boldsymbol{\nu}}_{\text{glom.} i}, \boldsymbol{\nu}_{\text{glom.} j}) \, ,
\end{equation}
with $d^{\text{Pearson}}(\boldsymbol{\cdot},\boldsymbol{\cdot})$ the Pearson correlation coefficient between two vectors.
For the input correlation matrix $\mathbf{C}^{\text{input}}$, the vector $\boldsymbol{\nu}_{\text{glom.}i}$ contained the average firing rates of the six RNs projecting to the $i$th glomerulus, 
with each element of this vector for one stimulus presentation. 
For the output correlation matrix $\mathbf{C}^{\text{output}}$ we used the rates from PNs instead of RNs.
Thus, we obtained $10 \times 10$ matrices containing the rate correlations for each pair of input or output channels. 

\prettyref{fig:insectAL}B depicts the correlation matrix $\mathbf{C}^{\text{input}}$ for the input firing rates.
When no lateral inhibition is present, $\mathbf{C}^{\text{input}}$ matches $\mathbf{C}^{\text{output}}$ (\prettyref{fig:insectAL}C). 
We have systematically varied the strength of lateral inhibition by scaling all inhibitory weights by a factor $q$, with $q=0$ for zero lateral inhibition and $q=1$ for inhibition set to its maximal strength.
With increasing lateral inhibition, off-diagonal values in $\mathbf{C}^{\text{output}}$ approach zero and output channel correlation is virtually gone (\prettyref{fig:insectAL}D).
The amount of residual correlation to be present in the output can be controlled by adjusting the strength of lateral inhibition (\prettyref{fig:insectAL}E).

Taken together, we demonstrated the implementation of an olfaction-inspired network to remove correlation between input channels on the \spikey{} chip. 
This network can serve as a preprocessing module for data analysis applications to be implemented on the \spikey{} chip. 
An interesting candidate for such an application is a spiking network for supervised classification, which may benefit strongly from reduced channel correlations for faster learning and better discrimination \citep{Hausler2011}. 
\subsection{Liquid State Machine}
\label{sec:lsm}

\emph{Liquid state machines} (LSMs) as proposed by \citet{Maass02_2531}
and \citet{Jaeger01_echo} provide a generic framework for computation on
continuous input streams. The \emph{liquid}, a recurrent network, projects an
input into a high-dimensional space which is subsequently read out.
It has been proven that LSMs have universal computational power for
computations with fading memory on functions of time \citep{Maass02_2531}.
In the following, we show that classification performance of an LSM emulated on
our hardware is comparable to the corresponding computer simulation.
Synaptic weights of the readout are iteratively learned on-chip, which inherently
compensates for fixed-pattern noise.
A trained system can then be used as an autonomous and very fast spiking
classifier.

\subsubsection{\network{}}

The LSM consists of two major components: the recurrent liquid network itself and a
spike-based classifier (\prettyref{fig:lsm}A).
A general purpose liquid needs to meet the separation property \citep{Maass02_2531}, which requires
that different inputs are mapped to different outputs, for a wide range of
possible inputs. Therefore, we use a network topology similar to the one proposed by
\citet{Bill10_129}.
It consists of an excitatory and inhibitory population with a ratio of 80:20 excitatory to inhibitory neurons. 
Both populations have recurrent as well as feedforward connections.
Each neuron in the liquid receives 4 inputs from the 32 excitatory and 32 inhibitory sources, respectively.
All other connection probabilities are illustrated in \prettyref{fig:lsm}.

The readout is realized by means of a tempotron \citep{Guetig06_420}, which is
compatible with our hardware due to its spike-based nature.
Furthermore, its modest single neuron implementation leaves most hardware
resources to the liquid. The afferent synaptic weights are trained with the
method described in \citet{Guetig06_420}, which effectively implements gradient
descent dynamics.
Upon training, the tempotron distinguishes between two input classes by
emitting either one or no spike within a certain time window. The former is
artificially enforced by blocking all further incoming spikes after the first
spike occurrence.

The PSP kernel of a LIF neuron with current-based
synapses is given by
\begin{equation}
    K(t-\itsub{t}{i}) =
    A \left(
        e^{-\frac{t-\itsub{t}{i}}{\taum}} -
        e^{-\frac{t-\itsub{t}{i}}{\tausyn}}
    \right) \cdot \Theta(t-\itsub{t}{i})
    \ ,
    \label{eq:tempotron_spike_contribution}
\end{equation}
with the membrane time constant $\taum$ and the synaptic time constant
$\tausyn$, respectively.
Here, $A$ denotes a constant PSP scaling factor, $\itsub{t}{i}$ the time of the $i$th
incoming spike and $\Theta(t)$ the Heaviside step function.

During learning, weights are updated as follows
\begin{equation}
    \Delta \itsub{\weight}{j}^n =
    \begin{cases}
        0 & \text{correct}\\
        \alpha(n) \sum_{\itsub{t}{i,j}<\mrmsub{t}{max}}
        K(\mrmsub{t}{max}-\itsub{t}{i,j}) &
            \text{erroneous},
    \end{cases}
    \label{eq:tempotron_learning_rule}
\end{equation}
where $\Delta \itsub{\weight}{j}^n$ is the weight update corresponding to the
$j$th afferent neuron after the $n$th learning iteration with learning rate
$\alpha(n)$. The spike time of the tempotron, or otherwise the time of highest membrane
potential, is denoted with $\mrmsub{t}{max}$.
In other words, for trials where an erroneous spike was elicited,
the excitatory afferents with a causal contribution to this spike
are weakened and inhibitory ones are strengthened according to \prettyref{eq:tempotron_learning_rule}.
In case the tempotron did not spike even though it should have, the
weights are modulated the other way round, i.e.\ excitatory weights are strengthened and inhibitory ones are weakened.
This learning rule has been implemented on hardware with small modifications, due to the
conductance-based nature of the hardware synapses (see below).

The tempotron is a binary classifier, hence any task needs to be mapped to a
set of binary decisions.
Here, we have chosen a simple binary task adapted from
\cite{Maass02_2531}, to evaluate the performance of the LSM. The challenge was
to distinguish spike train segments in a continuous
data stream composed of two templates with identical rates (denoted X and Y in \prettyref{fig:lsm}A).
In order to generate the input, we cut the template spike trains
into segments of \SI{50}{\milli\second} duration.
We then composed the spike sequence to be presented to the network by randomly picking a spike segment from either X or Y in each time window (see \prettyref{fig:lsm} for a schematic).
Additionally, we added spike timing jitter from a normal distribution with a standard deviation of $\sigma = \SI{1}{\milli\second}$ to each spike.
For each experiment run, both for training and evaluation, the composed spike sequence was then streamed into the liquid.
Tempotrons were given the liquid activity as input and trained to identify whether the segment within the previous time window originated from sequence X or Y. 
In a second attempt, we trained the tempotron to identify the origin of the pattern presented in the window at -$100$ to -\SI{150}{\milli\second} (that is, the second to the last window). 
Not only did this task allow to determine the classification capabilities of the LSM, but
it also put the liquid's fading memory to the test, as classification of a
segment further back in time becomes increasingly difficult.

\begin{figure*}[tb]
\ifdraftmode
  \mbox{
  \hspace{-2.25cm}
\fi
\includegraphics[width=\fullfigwidth]{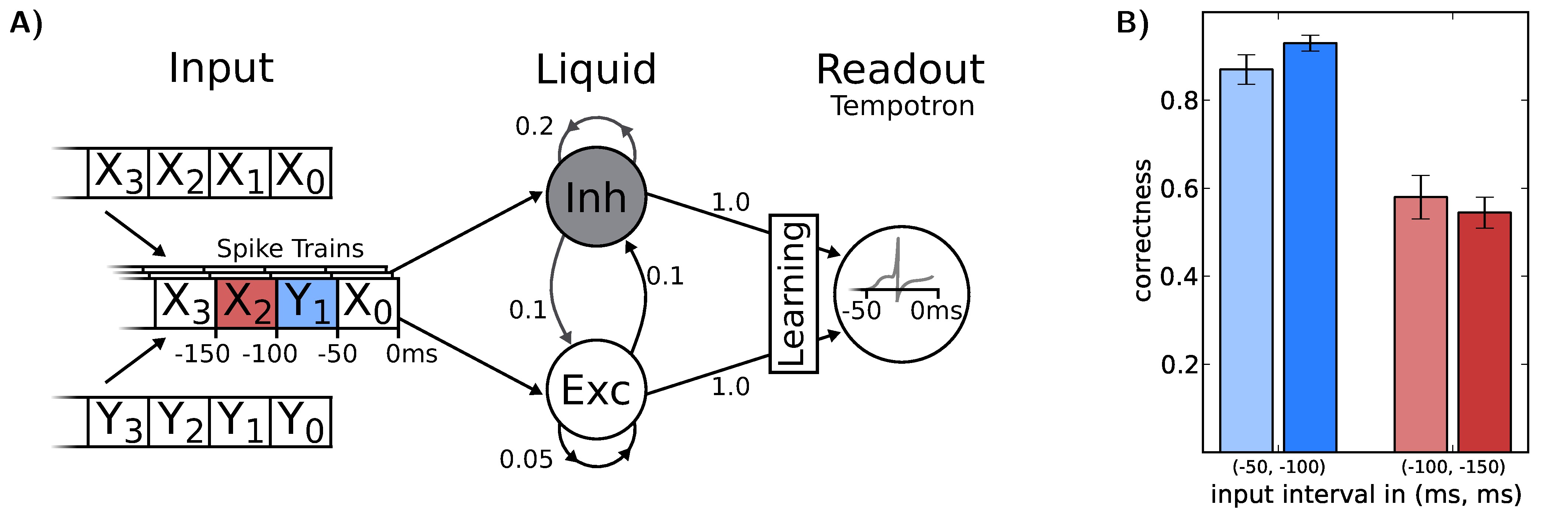}
\ifdraftmode
  }
\fi
\caption{\textbf{(A)}
Schematic of the LSM and the given task.
Spike sources are composed of \SI{50}{\milli\second} segments drawn from two template spike trains (X and Y).
These patterns are streamed into the liquid (with descending index), which is a network consisting of 191 neurons, leaving one neuron for the tempotron.
Connection probabilities are depicted next to each connection (arrows).
In two experiments, the tempotron is trained to either classify the origin (X or Y) of the spike train segment with index $1$ or $2$.
\textbf{(B)}
The classification performance of the LSM measured over 200 samples after 1000
training iterations for both hardware (lighter) and software (darker)
implementation.
\label{fig:lsm}}
\end{figure*}

\subsubsection{\hardware{}}

The liquid itself does not impose any strong requirements on the hardware since
virtually any network is suitable as long as the separation
property is satisfied.
We adapted a network from \citet{Bill10_129}
which, in a similar form, had already been implemented on our hardware.
However, STP was disabled,
because at the time of the experiment it was not possible to exclusively enable STP for the liquid without severely
affecting the performance of the tempotron.

The hardware implementation of the tempotron required more
attention, since only conductance-based synapses are available.
The dependence of spike efficacies on the actual membrane potential was
neglected, because the rest potential was
chosen to be close to the firing threshold, with the reversal potentials
far away.
However, the asymmetric distance of excitatory and inhibitory reversal
potentials from the sub-threshold regime needed compensation.
This was achieved by scaling all excitatory weights by
$(\overline{\vmem}-\einh)/(\overline{\vmem}-\eexc)$, where
$\overline{\vmem}$ corresponds to the mean neuron membrane voltage and
$\eexc$/$\einh$ is the excitatory/inhibitory reversal
potentials.
Discontinuities in spike efficacies for synapses changing from excitatory to
inhibitory or vice versa were avoided by prohibiting such transitions.
Finally, membrane potential shunting after the first spike occurrence is
neither possible on our hardware nor very biological and had therefore
been neglected, as already proposed by \cite{Guetig06_420}.

Even though the tempotron was robust against fixed-pattern noise due to on-chip learning, the liquid required modifications.
Therefore, firing thresholds were tuned independently in software and
hardware to optimize the memory capacity and avoid violations of the separation property.
Since hardware neurons share firing thresholds, the tempotron was affected
accordingly (see \prettyref{tab:paramlist}).
Additionally, the learning curve $\alpha(n)$ was chosen individually for
software and hardware due to the limited resolution of synaptic weights on the latter.

The results for software and hardware implementations are illustrated in
\prettyref{fig:lsm}B.
Both LSMs performed at around 90\% classification correctness for the
spiketrain segment that lied \SI{50}{\milli\second} to \SI{100}{\milli\second} in the past with respect to the
end of the stimulus.
For inputs lying even further away in time, performances dropped to chance
level (50\% for a binary task), independent of the simulation back-end.

Regarding the classification capabilities of the LSM, our current
implementation allows a large variety of tasks to be performed.
Currently, e.g., we are working on hand-written
digit recognition with the very
same setup on the \spikey{} chip. Even without a liquid, our implementation of the tempotron
(or populations thereof) makes an excellent neuromorphic
classifier, given its bandwidth-friendly sparse response and robustness
against fixed-pattern noise.
\section{Discussion}

We have successfully implemented a variety of neural microcircuits on a single universal neuromorphic substrate, which is described in detail by \citet{Schemmel06_1}.
All networks show activity patterns qualitatively and to some extent also quantitatively similar to those obtained by software simulations.
The corresponding reference models found in literature have not been modified significantly and network topologies have been identical for hardware emulation and software simulation, if not stated otherwise.
In particular, the emulations benefit from the advantages of our neuromorphic implementation, namely inherent parallelism and accelerated operation compared to software simulations on conventional von-Neumann machines.
Previous accounts of networks implemented on the \spikey{} system include computing with high-conductance states \citep{Kaplan09_1524}, self-stabilizing recurrent networks \citep{Bill10_129}, and simple emulations of cortical layer 2/3 attractor networks \citep{Bruederle11_263}.

In this contribution, we have presented a number of new networks and extensions of previous implementations.
Our synfire chain implementation achieves reliable signal propagation over years of biological time from one single stimulation, while synchronizing and filtering these signals (\prettyref{sec:sfc}).
Our extension of the network from \citet{Bill10_129} to exhibit asynchronous irregular firing behavior is an important achievement in the context of reproducing stochastic activity patterns found in cortex (\prettyref{sec:brn}). 
We have realized soft winner-take-all networks on our hardware system (\prettyref{sec:wta}), which are essential building blocks for many cortical models involving some kind of attractor states \citep[e.g., the decision-making model by][]{Soltani10_112}.
The emulated cortical attractor model provides an implementation of working memory for computation with cortical columns (\prettyref{sec:kth}).
Additionally, we have used the \spikey{} system for preprocessing of multivariate data inspired by biological archetypes (\prettyref{sec:al}) and machine learning (\prettyref{sec:lsm}).
Most of these networks allocate the full number of neurons receiving input from one synapse array on the \spikey{} chip, but with different sets of neuron and synapse parameters and especially vastly different connectivity patterns, thereby emphasizing the remarkable configurability of our neuromorphic substrate.

However, the translation of such models requires modifications to allow execution on our hardware. 
The most prominent cause for such modifications is fixed-pattern noise across analog hardware neurons and synapses.
In most cases, especially when population rate coding is involved, it is sufficient to compensate for this variability by averaging spiking activity over many neurons. 
For the data decorrelation and machine learning models, we have additionally trained the synaptic weights on the chip to achieve finer equilibration of the variability at critical network nodes.
Especially when massive downscaling is required in order for models to fit onto the substrate, fixed pattern noise presents an additional challenge because the same amount of information needs to be encoded by fewer units.
For this reason, the implementation of the cortical attractor memory network required additional heuristic activity fitting procedures.

The usability of the \spikey{} system, especially for neuroscientists with no neuromorphic engineering background, is provided by an integrated development environment.
We envision that the configurability made accessible by such a software environment will encourage a broader neuroscience community to use our hardware system.
Examples of use would be the acceleration of simulations as well as the investigation of the robustness of network models against parameter variability, both between computational units and between trials, as e.g.\ published by \citet{Bruederle10_iscas} and \citet{Schmuker11_109}.
The hardware system can be efficiently used without knowledge about the hardware implementation on transistor level.
Nevertheless, users have to consider basic hardware constraints, as e.g., shared parameters.
Networks can be developed using the PyNN metalanguage and optionally be prototyped on software simulators before running on the \spikey{} system \citep{Davison09, Bruederle09_17}.
This rather easy configuration and operation of the \spikey{} chip allows the implementation of many other neural network models.

There exist also boundaries to the universal applicability of our hardware system. 
One limitation inherent to this type of neuromorphic device is the choice of implemented models for neuron and synapse dynamics.
Models requiring, e.g., neuronal adaptation or exotic synaptic plasticity rules are difficult, if not impossible to be emulated on this substrate.
Also, the total number of neurons and synapses set a hard upper bound on the size of networks that can be emulated.
However, the next generation of our highly accelerated hardware system will increase the number of available neurons and synapses by a factor of $10^3$, and provide extended configurability for each of these units \citep{Schemmel10_1947}.

The main purpose of our hardware system is to provide a flexible platform for highly accelerated emulation of spiking neuronal networks.
Other research groups pursue different design goals for their hardware systems. 
Some focus on dedicated hardware providing specific network topologies \citep[e.g.,][]{Merolla06_4539,Chicca07_981}, or comprising few neurons with more complex dynamics \citep[e.g.,][]{Chen10_1511,Grassia11_134,Brink12_1}.
Others develop hardware systems of comparable configurability, but operate in biological real-time, mostly using off-chip communication \citep{Vogelstein07_253,Choudhary12_121}.
Purely digital systems \citep{Merolla11_01,Furber12_1,Imam12_25} and field-programmable analog arrays \citep[FPAA;][]{Basu10_311} provide even more flexibility in configuration than our system, but have much smaller acceleration factors.
\verysmalltodominor{TP: one may add SpiNNaker somewhere here}

With the ultimate goal of brain size emulations, there exists a clear requirement for increasing the size and complexity of neuromorphic substrates.
An accompanying upscaling of the fitting and calibration procedures presented here appears impractical for such orders of magnitude and can only be done for a small subset of components.
Rather, it will be essential to step beyond simulation equivalence as a quality criterion for neuromorphic computing, and to develop a theoretical framework for circuits that are robust against, or even exploit the inherent imperfections of the substrate for achieving the required computational functions.
\verysmalltodominor{TP: here we could also say: nobody knows how noise effects computation -> precise sw is not better than hw -> robust models are more promising}
\ifdraftmode\clearpage{}\fi
\section*{Acknowledgments}

We would like to thank Dan Husmann, Stefan Philipp, Bernhard Kaplan, and Moritz Schilling for their essential contributions to the neuromorphic platform,
Johannes Bill, Jens Kremkow, Anders Lansner, Mikael Lundqvist and Emre Neftci for assisting with the hardware implementation of the network models,
Oliver Breitwieser for data analysis and Venelin Petkov for characterisation measurements of the \spikey{} chip.

The research leading to these results has received funding by the European Union 6th and 7th Framework Programme under grant agreement no.\ 15879 (FACETS), no.\ 269921 (BrainScaleS) and no. 243914 (Brain-i-Nets).
Michael Schmuker has received support from the german ministry for research and education (BMBF) to Bernstein Center for Computational Neuroscience Berlin (grant no.\ 01GQ1001D) and from Deutsche Forschungsgemeinschaft within SPP 1392 (grant no.\ SCHM 2474/1-1).

The main contributors for each section are denoted with author initials:
neuromorphic system (AG \& EM);
synfire chain (PM);
balanced random network (TP);
soft winner-take-all network (TP);
cortical attractor model (MP);
antennal lobe model (MS);
liquid state machine (SJ).
All authors contributed to writing this paper.

\bibliographystyle{neuralcomput_natbib}
\small
\bibliography{bib}
\normalsize
\clearpage{}



\end{document}